\documentclass[11pt,a4paper]{article}
\pdfoutput=1
\usepackage{jheppub}
\usepackage{gensymb}
\usepackage{subcaption}
\usepackage{mwe}
\usepackage{amssymb,amsmath}
\usepackage{graphicx}
\usepackage{color}
\usepackage{cancel}
\usepackage{booktabs}
\usepackage[colorlinks=true
,urlcolor=blue
,citecolor=blue
,linkcolor=blue
,pagecolor=blue
,linktocpage=true
,pdfproducer=medialab
]{hyperref}

\usepackage[numbers]{natbib}
\usepackage{notoccite}
\usepackage{comment} 
\usepackage{placeins}
\usepackage{bigints}
\makeatletter \renewcommand{\@dotsep}{10000} \makeatother
\def\be{\begin{equation}}
\def\ee{\end{equation}}
\def\bea{\begin{eqnarray}}
\def\eea{\end{eqnarray}}
\def\bi{\begin{itemize}}
\def\ei{\end{itemize}}

\def\to{\rightarrow}

\usepackage[nodisplayskipstretch]{setspace}

\def\mg2{muon $g-2$}
\newcommand{\cred}[1]{{\bf \color{red} #1}}
\newcommand{\cblue}[1]{{\bf \color{blue} #1}}
\DeclareUnicodeCharacter{2212}{-}
\DeclareUnicodeCharacter{202F}{\,}

\begin{document}

\begin{titlepage}
\pagestyle{empty}

\vspace*{0.2in}
\begin{center}
{\Large \bf Non-holomorphic Contributions in GMSB with  \\ Adjoint Messengers}

\vspace{1cm}
{\bf B\"{u}\c{s}ra Ni\c{s}\footnote{E-mail: 501507008@ogr.uludag.edu.tr} and Cem Salih $\ddot{\rm U}$n\footnote{E-mail: cemsalihun@uludag.edu.tr}}
\vspace{0.5cm}

{\small \it Department of Physics, Bursa Uluda\~{g} University, TR16059 Bursa, Turkey}

\end{center}

\vspace{0.5cm}
\begin{abstract}

We consider models of gauge mediated supersymmetry breaking, in which the breaking is transmitted to the visible sector by the messenger fields from the adjoint representation of MSSM's gauge group. In addition, we include the non-holomorphic terms induced by the supersymmetry breaking and involve them in the renormalization group evolution of the soft supersymmetry breaking terms. The main impact from the non-holomorphic terms arises in the right-handed stau mass, which requires large hypercharge interactions with the messengers to accommodate non-tachyonic staus. With the non-holomorphic contributions, the stau mass-square can be driven to the positive values in the renormalization group evolution, even if the hypercharge interactions are small. Even though the radiative non-holomorphic contributions enhance the mass spectrum, their effects in the sparticle mixing rather reduce their overall contributions such that we realize about 25 GeV difference in the right-handed stau mass, while the difference is lowered by about 5 TeV in the lightest mass-eigenstate of staus. In addition to staus, we realize 6-7 TeV difference in the sbottom mass, and about 15 TeV in the stop mass when the non-holomorphic terms are largely induced. These contributions in the sparticle masses also affect the SM-like Higgs boson mass, and we find that the SM-like Higgs boson can be enhanced as much as about 80 GeV. In a small region of the parameter space we also observe negative non-holomorphic contributions which do not exceed about 1 TeV for the sparticles, and 20 GeV for the SM-like Higgs boson. An interesting impact from the non-holomorphic terms happens in the muon $g-2$ results. Its latest experimental and theoretical results may remove need for new physics, and hence one can formulate further suppressions in the supersymmetric contributions to muon $g-2$. We find that the non-holomorphic contributions can provide a significant decreasing in supersymmetric contributions to muon $g-2$, and even a large muon $g-2$ results can be fit with the current bound after adding the non-holomorphic contributions. We realize that the muon $g-2$ results can be decreased as much as about $-50\times 10^{-10}$  by the non-holomorphic contributions, and consequently one can still accommodate light sleptons and gauginos in the spectrum.

\end{abstract}
\end{titlepage}


\section{Introduction}
\label{sec:intro}

In addition to the completing the Standard Model (SM), the breakthrough discovery of the Higgs boson \cite{ATLAS:2012yve,CMS:2012qbp,CMS:2013btf} is not only an observation of new particle, but it also defines and shape the path through the models beyond the SM (BSM). While the long standing problems of SM such as the gauge hierarchy problem, absence of dark matter (DM) have already led to the need for BSM, the detailed analyses on the Higgs boson have pointed out also new problems as the stability of the Higgs potential within SM \cite{Degrassi:2012ry,Bezrukov:2012sa,Buttazzo:2013uya}. These analyses have yielded a strong impact on BSM studies, as well. Even the minimal supersymmetric extension of SM (MSSM), as being one of the appealing candidates for BSM by proposing a solution to the hierarchy problem, needs to heavy superysmmetric particles to accommodate a consistent Higgs boson in the mass spectrum \cite{Carena:2012mw,Carena:2011aa}. Although this requirement on supersymmetric (SUSY) spectra directly affects the stops, its impact spreads over the whole spectrum when SUSY models are constrained at a high scale at which the SUSY breaking is transmitted to the visible sector through flavour blind interactions as proposed by models of gravity mediated or gauge mediated SUSY breaking (GMSB) scenarios. In these scenarios, the SUSY breaking induces universal soft SUSY breaking (SSB) mass terms for all the three families, and the need for heavy stops keeps the other SUSY particles heavy, even though they do not contribute much to the Higgs boson mass. This impact becomes even severer in GMSB scenarios, since the mixing in the stop sector is negligible, whose contributions to the Higgs boson mass are compensated with quite heavy stops ($m_{\tilde{t}} = \mathcal{O}(10)$ TeV) \cite{Ajaib:2012vc,Gogoladze:2015tfa}.

The level of the sensitivity in the current experimental analyses in the Higgs boson mass ($\Delta m_{h} = \mathcal{O}(100)$ MeV) \cite{ParticleDataGroup:2024cfk} requires also more sensitive calculations, which might be accomplished by matching conditions at different energy scales, fixed-point calculations \cite{Goodsell:2014bna} and/or considering radiative corrections from the higher order renormalization group equations (RGEs) \cite{Allanach:2001kg,Allanach:2014nba}. Such improvements can shift the theoretical results of the Higgs boson mass up to about 3 GeV when the mixing among the stops is large \cite{Baer:2021tta}. On the other hand, in the models such as those in GMSB class, this shift cannot exceed about 0.5 GeV \cite{Gomez:2022qrb}, since the mixing in stops is negligible in this class. In this context, despite being necessary, such contributions are not large enough to ameliorate the impact from the Higgs boson mass on the mass spectrum. Thus, one can also consider to extend the symmetry \cite{Gogoladze:2015jua} and/or several messengers which carry SUSY breaking to the visible sector.

In our work, we consider a class of GMSB models which propose several messenger fields transmitting the SUSY breaking to the visible sector through the SM gauge interactions. One can construct such models with the messengers from the adjoint representation of the SM gauge group \cite{Han:1998pa,Bhattacharyya:2013xma}. An incomplete version of these models allows to accommodate relatively light weakly interacting SUSY particles in the spectra, by splitting the sleptons from squarks, and it can lead to interesting implications which can be tested experimentally \cite{Bhattacharyya:2015vha, Gogoladze:2016grr,Gogoladze:2016jvm}. However, this version suffers from inconsistently light sleptons and needs to be extended to fit consistent masses for the charged sleptons \cite{Gogoladze:2016grr,Gogoladze:2016jvm}. 

In addition to consider the full representation for the messenger fields, we also invoke the contributions from some non-holomorphic (NH) terms in our analyses. Such terms are strictly forbidden by SUSY due to the holomorphy condition; however, they can be induced during the SUSY breaking \cite{Inoue:1982pi,Hall:1990ac,Bagger:1993ji,Bagger:1995ay,Martin:1999hc,Jack:1999ud,Jack:1999fa,Haber:2007dj}. The emergence of such terms might be a sign of hard SUSY breaking, and they can lead to the hierarchy problem if SUSY is broken through a VEV of a field which is singlet under MSSM gauge group. However, they can be listed as the SSB terms in the absence of a MSSM singlet field \cite{Inoue:1982pi,Martin:1999hc,Jack:1999ud,Jack:1999fa}. Due to the holomorphy condition before the SUSY breaking, one needs to consider one loop further in the perturbation theory \cite{Haber:2007dj}, and the dimensionality indicates a further suppression by $1/M$ in NH terms compared with the usual holomorphic SSB terms \cite{Bagger:1993ji,Martin:1999hc}. This suppression might be considered as to be the reason that these terms have been neglected mostly, but some recent studies have shown that the impact from these parameters can be considerable even when they are small at the SUSY breaking scale \cite{Chattopadhyay:2017qvh,Ali:2021kxa,Chakraborty:2019wav,Un:2014afa,Rehman:2022ydc,Un:2023wws,Israr:2025cfd,Israr:2024ubp,Rehman:2025djc}.

On the other hand, previous studies mostly consider the NH terms effectively by including them at the decoupling scale ($M_{{\rm SUSY}}\equiv \sqrt{m_{t_{L}}m_{\tilde{t}_{R}}}$, where $m_{\tilde{t}_{L}}$ and $m_{\tilde{t}_{R}}$ are the SSB masses of the left and right-handed stops, respectively). In these cases, the SSB terms evolve from a high scale to the electroweak scale without NH terms, and their effects rather emerge from the mixing of the sfermions and their couplings to the ``wrong" Higgs field \cite{Haber:2007dj}. Even though they can significantly alter the low scale phenomenology through their contributions to the masses of sfermions and the SM-like Higgs boson, such effects might be cancelled out if the SSB parameters and couplings evolve from a high scale with the NH terms through renormalization group equations (RGEs). In our work, we consider the NH contributions by including these terms in RGEs by linking them to the SUSY breaking, which leads to NH terms suppressed by $1/M$. The rest of the paper is organized as follows: We describe the models with the adjoint messengers in GMSB models (hereafter, GMSB-ADJ) by considering the induced SSB terms in Section \ref{sec:model}. We also introduce the NH terms and their contributions to the observables in this section. Section \ref{sec:scan} briefly summarize the data generation and experimental constraints employed in our analyses. We discuss our results for the fundamental parameter space of the models, the NH contributions to the mass spectra including the Higgs boson in Section \ref{sec:fund}. After the latest release about the experimental and theoretical results of muon anomalous magnetic moment (muon $g-2$), we also add a note about the NH impact on muon $g-2$ in Section \ref{sec:muong2}. Finally we summarize and conclude our results in Section \ref{sec:conc}.

\section{SSB with the Adjoint Messengers}
\label{sec:model}

In the GMSB models, the SUSY breaking is induced in the visible sector through the gauge interactions of the MSSM gauge group. The minimal models of this class usually consider the messengers resided in $\mathbf{5} + \mathbf{\bar{5}}$ representations of $SU(5)$ and/or $\mathbf{10} + \mathbf{\bar{10}}$ of $SO(10)$. Despite their nice flavour blind structures, these minimal models are those which feel the strongest negative impact from the Higgs boson discovery \cite{Ajaib:2012vc,Gogoladze:2015tfa,Gogoladze:2015jua,Hamaguchi:2014sea,Delgado:2013gza,Draper:2011aa,Draper:2013oza}. Even though it is slightly away from the minimality, one can also consider several messengers resided in the adjoint representation of the MSSM gauge group. In these models of the GMSB-ADJ class, the messengers interact with a gauge singlet field $\hat{S} = S + \theta^{\alpha}\psi_{\alpha} + \theta^{2}F$  in the hidden sector through the following superpotential:

\begin{equation}
W_{{\rm ADJ}} = (\mathcal{M}_{5} + \lambda_{5}\hat{S})\Sigma_{5}\Sigma_{5} +  (\mathcal{M}_{3} + \lambda_{3}\hat{S}){\rm TR}(\Sigma_{3}\Sigma_{3}) + (\mathcal{M}_{8} + \lambda_{8}\hat{S}){\rm TR}(\Sigma_{8}\Sigma_{8})~,
\label{eq:messengersuperpotential}
\end{equation} 
where $\Sigma_{5}$, $\Sigma_{3}$ and $\Sigma_{8}$ represent the messenger fields which interact with the MSSM fields through $U(1)_{Y}$, $SU(2)_{L}$ and $SU(3)_{C}$, respectively. Note that $\Sigma_{5}$ in our notation corresponds to the fields in $\mathbf{5}+\mathbf{\bar{5}}$. Besides, the gauge singlet field $\hat{S}$ in this superpotential is not one of those taking part in hard breaking of SUSY, since it does not interact with the MSSM fields directly. The terms $\mathcal{M}_{5}$, $\mathcal{M}_{3}$ and $\mathcal{M}_{8}$ stand for masses of the messenger fields, and, for simplicity, they are set to be equal in our analyses as $\mathcal{M}_{5}=\mathcal{M}_{3}=\mathcal{M}_{8} \equiv M_{{\rm Mess}}$, which is expected to be at the order of $\langle S \rangle$. SUSY breaking VEV of $\hat{S}$ can be expressed in terms of its scalar components as $\langle \hat{S} \rangle  = \langle S \rangle + \theta^{2}\langle F \rangle$, which induces different masses for the bosonic and fermionic components of the messenger fields as 

\begin{equation}
m_{b_{i}} = M_{{\rm Mess}}\sqrt{1\pm \dfrac{\Lambda_{i}}{M_{{\rm Mess}}}}~, m_{f_{i}} = M_{{\rm Mess}}~,\hspace{0.3cm} i = 5,3,8~;
\label{eq:messengerMass}
\end{equation}
where $\Lambda \equiv \lambda_{i}\langle F \rangle/M_{{\rm Mess}}$. As seen from the mass expression of the bosonic component, the condition $\Lambda_{i} < M_{{\rm Mess}}$ should hold to avoid tachyonic states ($m_{b_{i}}^{2} < 0 $). The SSB masses for the gauginos and the matter fields emerge in the same way as described for the minimal GMSB models: i.e. the SSB gaugino masses at one-loop and at two-loop for the scalars \cite{Giudice:1998bp}. In the case of the adjoint messengers, the SSB masses can be derived \cite{Bhattacharyya:2013xma} for the gauginos as

\begin{equation*}
M_1 \simeq \frac{g_1^2}{16 \pi^2} (5 \Lambda_5)~,\hspace{0.5cm}
M_2 \simeq \frac{g_2^2}{16 \pi^2} (2 \Lambda_3 + 3 \Lambda_5)~,\hspace{0.5cm}
M_3 \simeq \frac{g_3^2}{16 \pi^2} (3 \Lambda_8 + 2 \Lambda_5)~, \tag{2.2-a}
\label{eq:massgaugionos}
\end{equation*}
and for the sfermions
\begin{equation*}
\setstretch{2.8}
\begin{array}{rl}
m_Q^2  & \simeq \dfrac{2}{(16 \pi^2)^2} \left[ \dfrac{4}{3} g_3^4 (3\Lambda_8^2 + 2 \Lambda_5^2) + \dfrac{3}{4} g_2^4 (2 \Lambda_3^2 + 3 \Lambda_5^2) + \dfrac{3}{5} g_1^4 (5 \Lambda_5^2) \dfrac{1}{6^2}  \right]
\\
m_{\tilde{U}}^2 & \simeq \dfrac{2}{(16 \pi^2)^2} \left[ \dfrac{4}{3} g_3^4 (3\Lambda_8^2 + 2 \Lambda_5^2) + \dfrac{3}{5} g_1^4 (5 \Lambda_5^2) \left( \dfrac{2}{3} \right)^2 \right]
\\
m_{\tilde{D}}^2 &\simeq \dfrac{2}{(16 \pi^2)^2} \left[ \dfrac{4}{3} g_3^4 (3\Lambda_8^2 + 2 \Lambda_5^2) + \dfrac{3}{5} g_1^4 (5 \Lambda_5^2) \dfrac{2}{3^2} \right]
\\
m_{\tilde{L}}^2 &\simeq \dfrac{2}{(16 \pi^2)^2} \left[ \dfrac{3}{4} g_2^4 (2\Lambda_3^2 + 3 \Lambda_5^2) + \dfrac{3}{5} g_1^4 (5 \Lambda_5^2) \dfrac{1}{2^2} \right];\hspace{0.5cm} m_{H_u}^2=m_{H_d}^2 =m_{\tilde{L}}^2 
\\
m_{\tilde{E}}^2 & \simeq \dfrac{2}{(16 \pi^2)^2} \left[ \dfrac{3}{5} g_1^4 (5 \Lambda_5^2) \right]~.
\\
\end{array} \tag{2.2-b}
\label{eq:sfermionmasses}
\end{equation*}
\setcounter{equation}{2}

As happens in the GMSB models, the trilinear SSB couplings can be induced at the third loop and so they are suppressed by an extra $M_{{\rm Mess}}$ factor. Thus they can be set to zero at the messenger scale safely. However, despite being negligibly small at $M_{{\rm Mess}}$, they can grow through RGEs easily up to about an order of TeV, and still be effective at the low scale observables. Similarly, the NH terms, in general, can be induced by the higher order operators effectively as \cite{Martin:1999hc,Bagger:1995ay}

\begin{equation}
\setstretch{2.5}
\begin{array}{lll}
\dfrac{1}{M^{3}}\bigint d^{4}x d^{4}\theta \hat{X}\hat{\bar{X}}^{\dagger}\hat{\Phi}_{i}\hat{\Phi}_{j}\hat{\Phi}_{k}^{\dagger} &  = \dfrac{\mathcal{F}^{2}}{M^{3}}S_{i}S_{j}S_{k}^{*} & \rightarrow A^{\prime}S_{i}S_{j}S_{k}^{*}~, \\
\dfrac{1}{M^{3}}\bigint d^{4}x d^{4}\theta \hat{X}\hat{\bar{X}}^{\dagger}\mathcal{D}^{\alpha}\hat{\Phi}\mathcal{D}_{\alpha}\hat{\Phi} & = \dfrac{\mathcal{F}^{2}}{M^{3}}\psi\psi & \rightarrow \mu^{\prime}\psi\psi~,
\end{array}
\label{eq:NH}
\end{equation}
which can be adapted to the GMSB-ADJ models by setting $M\rightarrow M_{{\rm Mess}}$ and $\mathcal{F}^{2}\rightarrow \displaystyle{\sum_{i}}\langle F_{i} \rangle^{2}$, and the typical order for the NH terms is found as 

\begin{equation}
A^{\prime}, \mu^{\prime} \sim \dfrac{1}{M_{{\rm Mess}}}\displaystyle{\sum_{i}}\Lambda_{i}^{2}~.
\label{eq:NHmess}
\end{equation}

In this context, one can also consider to set the NH terms to zero as can be done for their holomorphic partners; however, it may not be a good approximation for the NH terms, since the RGEs do not have a growing effect on the NH terms unless one assumes another source which induces considerable NH interactions \cite{McGuirk:2012sb}, which can be seen from the following RGEs of the NH terms:

\begin{equation}\hspace{-0.8cm}
\setstretch{2.5}
\begin{array}{rl}
\dfrac{d\mu^{\prime}}{dt} & = \dfrac{1}{16\pi^{2}}\left(y_{\tau}^{2} + 3y_{b}^{2}+3y_{t}^{2} -3g_{2}^{2} + \dfrac{3}{5}g_{1}^{2} \right)\mu^{\prime} \\
\dfrac{d A^{\prime}_{t}}{dt} & = \dfrac{1}{16\pi^{2}}\left[\left(y_{\tau}^{2} + 5y_{b}^{2} + 3y_{t}^{2} \right)A^{\prime}_{t} + 2y_{b}^{2}\left(A^{\prime}_{b} -2\mu^{\prime}\right) + \left(A^{\prime}_{t} - 2\mu^{\prime}\right)\left(3g_{2}^{2} + \dfrac{3}{5}g_{1}^{2}\right)\right] \\
\dfrac{dA_{b}^{\prime}}{dt} & = \dfrac{1}{16\pi^{2}}\left[ \left(-y_{\tau}^{2} + 3y_{b}^{2} + 5y_{t}^{2} \right)A_{b}^{\prime} + 2y_{t}^{2}\left(A^{\prime}_{t} - 2\mu^{\prime}\right) + \left(A_{b}^{\prime} - 2\mu^{\prime}\right)\left(3g_{2}^{2} + \dfrac{3}{5}g_{1}^{2}\right)\right] \\
\dfrac{d A_{\tau}^{\prime}}{dt} & = \dfrac{1}{16\pi^{2}}\left[ \left(y_{\tau}^{2} -3y_{b}^{2} + 3y_{t}^{2} \right)A_{\tau}^{\prime} + 6y_{b}^{2}A_{b}^{\prime} +\left(A_{\tau}^{\prime} - 2\mu^{\prime}\right)\left(3g_{2}^{2} + \dfrac{3}{5}g_{1}^{2}\right)\right]
\end{array}
\label{eq:NHRGE}
\end{equation}

As is seen from the RGEs given in Eq.(\ref{eq:NHRGE}), setting the NH terms to zero at $M_{{\rm Mess}}$ fixes them to zero at all the energies. Hence, we keep them small but non-zero in our analyses. The resultant SSB Lagrangian, then, can be summarized as follows:

\begin{equation}
\setstretch{2.0}
\begin{array}{ll}
\mathcal{L}^{{\rm MSSM}}_{{\rm soft}} = & -\dfrac{1}{2}\left(M_{1}\tilde{B}\tilde{B} + M_{2}\tilde{W}\tilde{W} + M_{3}\tilde{g}\tilde{g} \right) \\
& - \left(\tilde{Q}^{\dagger}m_{Q}^{2}\tilde{Q} + \tilde{L}^{\dagger}m_{L}^{2}\tilde{L} + \tilde{\bar{u}}m_{u}^{2}\tilde{\bar{u}}^{\dagger} + \tilde{\bar{d}}m_{d}^{2}\tilde{\bar{d}}^{\dagger} + \tilde{\bar{e}}m_{e}^{2}\tilde{\bar{e}}^{\dagger}\right) \\
& -m_{H_{u}}^{2}H_{u}^{\dagger}H_{u} -m_{H_{d}}^{2}H_{d}^{\dagger}H_{d} \\
& -(A_{u}\bar{Q}H_{u}u + A_{d}\bar{Q}H_{d}d + A_{e}\bar{L}H_{d}e)  \\
& -\mu^\prime {\tilde H_u}\cdot
{\tilde H_d} -\tilde{Q}~{H}_d^{\dagger} A^\prime_{u} \tilde{U}-
\tilde{Q}~ {H}_u^{\dagger} A^\prime_{d} \tilde{D}
 - \tilde{L}~{H}_u^{\dagger} A^\prime_{e} \tilde{E} - \mbox{h.c.}
\end{array}
\label{eq:SSBLag}
\end{equation}

In the MSSM, as mentioned before, the Higgs boson mass receives radiative contributions from stops and their mixing, while those from sbottom and stau can provide minor improvements, especially in the moderate $\tan\beta$ region \cite{Carena:2012mw}. In the presence of the NH terms, such minor contributions can be enhanced slightly \cite{Rehman:2022ydc,Un:2023wws}. In addition, the NH terms impose direct couplings for sbottom and stau to $H_{u}$, and such terms allow these supersymmetric particles to contribute directly to the SM-like Higgs boson, since SM-like Higgs boson is formed mostly by $H_{u}$ because of its direct coupling to the top quarks \cite{Gogoladze:2012ii}. Such contributions can diminish the impact from the Higgs boson analyses on the implications of BSM models.

\subsection{NH Contributions to SUSY Spectra} 
\label{subsec:SUSYMass}

The main radiative supersymmetric contributions to the Higgs boson mass arise from the self-energy diagrams in which the squarks and sleptons run in the loop. In this context the mixing in the mass-square matrices of these supersymmetric particles plays a crucial role. Indeed,  the cancellation between the fermion and sfermion contributions cannot hold after the SUSY breaking, and the trilinear couplings, which appear in the mixing, can be fit to realize finite and positive radiative contributions to the Higgs boson mass. The NH terms directly alter the mixing among these particles, and their effects can be understood best by considering the relevant mass-square matrices

\begin{equation}
 m_{\tilde{f}}^{2} = \left( \begin{array}{cc}
 m_{\tilde{f}_{L}\tilde{f}^{*}_{L}} & X_{\tilde{f}} \\ \\
X_{\tilde{f}}^{*} & m_{\tilde{f}_{R}\tilde{f}^{*}_{R}}
 \end{array} \right)
 \label{sfermionsmass2}
 \end{equation}
in the basis $(\tilde{f}_{L}, \tilde{f}_{R})$ and $(\tilde{f}^{*}_{L}, \tilde{f}^{*}_{R})$. The diagonal elements in the mass-square matrix remain the same as those in the MSSM \cite{Martin:1997ns}, the mixing terms are modified by the NH terms as follows:

\begin{equation*}
X_{\tilde{t}} = -\dfrac{m_{t}}{\sqrt{2}}\left[(\mu +A^{\prime}_{t})\cot\beta- A_{t}\right], \tag{2.8-a}
\label{eq:Xu}
\end{equation*}
\begin{equation*}
X_{\tilde{b}} = -\dfrac{m_{b}}{\sqrt{2}}\left[(\mu +A^{\prime}_{b})\tan\beta-A_{b}\right], \tag{2.8-b}
\label{eq:Xd}
\end{equation*}
\begin{equation*}
X_{\tilde{\tau}} = -\dfrac{m_{\tau}}{\sqrt{2}}\left[(\mu +A^{\prime}_{\tau})\tan\beta+A_{\tau}\right]. \tag{2.8-c}
\label{eq:Xe}
\end{equation*}
\setcounter{equation}{8}

The contributions from the NH term associated with the stops ($A^{\prime}_{t}$) are inversely proportional to $\tan\beta$, and hence, one can expect some minor contributions to the Higgs boson mass compared to those from its holomorphic partner, $A_{t}$. On the other hand, these terms appear together with $\tan\beta$ in the sbottom and stau mass-square matrices, and their contributions to the Higgs boson mass are enhanced with $\tan\beta$. We should note about the stability of the scalar potential before concluding this discussion that the coupling of the Higgs boson to staus, in particular, may lead to instability \cite{Carena:2012mw,Kitahara:2013lfa}. 

Once the NH terms are taken into account effectively at the low scale, their contributions to the masses appears only in their mixing, as summarized above, while the diagonal elements remain intact in the mass-square matrices given in Eq.(\ref{eq:sfermionmasses}). On the other hand, if the NH terms are imposed at the SUSY breaking scale in the visible sector ($M_{{\rm Mess}}$ in GMSB models), and involved in the RGEs, then the mass-squares of the sleptons and squarks result in different masses in the flavour basis after their RG evolutions. The NH effects in these mass terms can be seen in the following RGEs:

\begin{equation}
\setstretch{3.0}
\begin{array}{lll}
\dfrac{d m_{\tilde{t}_{R}}^{2}}{dt} & = \left(\dfrac{dm_{\tilde{t}_{R}}^{2}}{dt}\right)_{{\rm MSSM}} & + 4y_{t}^{2}(A^{\prime}_{t} - 2|\mu^{\prime}|^{2}) \\
\dfrac{d m_{\tilde{b}_{R}}^{2}}{dt} & = \left(\dfrac{dm_{\tilde{b}_{R}}^{2}}{dt}\right)_{{\rm MSSM}} & + 4y_{b}^{2}(A^{\prime}_{t} - 2|\mu^{\prime}|^{2}) \\
\dfrac{d m_{\tilde{t}_{L}}^{2}}{dt} & = \left(\dfrac{dm_{\tilde{t}_{L}}^{2}}{dt}\right)_{{\rm MSSM}} & + 2y_{t}^{2}A^{\prime}_{t} + 2y_{b}^{2}A^{\prime}_{b} - 4|\mu^{\prime}|^{2}(y_{t}^{2} + y_{b}^{2}) \\ 
\dfrac{d m_{\tilde{\tau}_{R}}^{2}}{dt} & = \left(\dfrac{dm_{\tilde{\tau}_{R}}^{2}}{dt}\right)_{{\rm MSSM}} & + 2y_{\tau}^{2}(A^{\prime}_{\tau} - 2|\mu^{\prime}|^{2}) \\
\dfrac{d m_{\tilde{\tau}_{L}}^{2}}{dt} & = \left(\dfrac{dm_{\tilde{\tau}_{L}}^{2}}{dt}\right)_{{\rm MSSM}} & + 2y_{\tau}^{2}(A^{\prime}_{\tau} - 2|\mu^{\prime}|^{2}) \\
\end{array}
\label{eq:NHMassRGEs}
\end{equation}

In this context, the NH terms can alter the masses in the full basis, and they even change the decoupling scale for the SUSY particles. Among the RGEs given above, the NH contributions can be crucial especially for the right-handed stau mass. As seen from the mass-square relations given in Eq.(\ref{eq:sfermionmasses}), small $\Lambda_{5}$ values can lead to inconsistently light or even tachyonic states for the right-handed stau field, which is one of the main factor shaping the fundamental parameter space of the models in the GMSB-ADJ class \cite{Gogoladze:2016grr,Gogoladze:2016jvm}. However, in the presence of the NH contributions, sufficiently large NH terms can keep the right-handed stau mass-square in consistent ranges. In addition to stau masses, the NH contributions to the stop masses can indirectly spread the NH effects over the whole spectra, since they change the SUSY decoupling scale $M_{{\rm SUSY}}$. In this context, even some particles, such as gauginos, do not receive any direct contribution from the NH terms, they will be affected since the NH terms will modify $M_{{\rm SUSY}}$ and so RGE running for all the SSB parameters.

\subsection{NH Contributions to the Higgs Boson Mass} 
\label{subsec:HiggsMass}

As is a well-known fact, the SM-like Higgs boson mass in MSSM is strongly based on its interactions with the stop. Hence, the mixing and masses in the stop sector play the most crucial role. The radiative contributions to the SM-like Higgs boson mass can be written as \cite{Carena:2012mw} 

\begin{equation*}
\Delta m_{h}^{2}\simeq \dfrac{m_{t}^{4}}{16\pi^{2}v^{2}\sin^{2}\beta}\dfrac{\mu A_{t}}{M^{2}_{{\rm SUSY}}}\left[\dfrac{A_{t}^{2}}{M^{2}_{{\rm SUSY}}}-6 \right]+
\end{equation*}
\begin{equation}\hspace{1.4cm}
\dfrac{y_{b}^{4}v^{2}}{16\pi^{2}}\sin^{2}\beta\dfrac{\mu^{3}A_{b}}{M^{4}_{{\rm SUSY}}}+\dfrac{y_{\tau}^{4}v^{2}}{48\pi^{2}}\sin^{2}\beta \dfrac{\mu^{3}A_{\tau}}{m_{\tilde{\tau}}^{4}}~~,
\label{eq:higgscor}
\end{equation}
where the first line represents the dominant contributions from stops, while the second line displays the minor contributions from sbottom and stau. These contributions in MSSM are restricted further by the stability condition of the Higgs potential \cite{Carena:2012mw,Kitahara:2013lfa}. In the presence of the NH terms, the trilinear couplings in Eq.(\ref{eq:higgscor}) should be replaced as $A_{i} \rightarrow X_{\tilde{i}}$, where $X_{\tilde{i}}$ are given in Eqs.(\ref{eq:Xu}, \ref{eq:Xd} and \ref{eq:Xe}) for $i=\tilde{t},\tilde{b},\tilde{\tau}$. respectively. With this inclusion of the NH terms, the contributions from stops are enhanced, but such an enhancements is expected to be small, since the NH contributions is suppressed by $\tan\beta$, and even it can be neglected compared to those from $A_{t}$. Despite small NH effects in stops through their mixing, they can still be significant with their RGE effects on the stop masses. In addition, the NH terms also induce a direct coupling between $\tilde{b},\tilde{\tau}$ to $H_{u}$, and they contribute to the SM-like Higgs boson directly as stop (for a detailed analyses see, for instance, \cite{Rehman:2022ydc,Rehman:2025djc}). 

\section{Scanning Procedure and Experimental Constraints}
\label{sec:scan}

In this section, we will briefly summarize the fundamental parameter space of the models in the GMSB-ADJ class, which is supplemented with the NH terms. Indeed, the NH terms do not bring more free parameters, unless one assumes different sources which can induce the NH terms through different mechanisms \cite{Bellisai:1997ck,Bergamin:2003ub,McGuirk:2012sb,Chattopadhyay:2017qvh}. In our work, we do not assume such mechanisms, and the NH terms are imposed at the messenger scale as given in Eq.(\ref{eq:NHmess}). Thus the fundamental parameters and their ranges in our analyses can be summarized as follows:

\begin{equation}
\begin{array}{rcl}
10^{-7}  \leq & {\Lambda}_{5, 3, 8} & \leq 10^{7}  ~~~{\rm GeV}\\
10^{6}~ \leq & M_{Mess}& \leq 10^{16}  ~~{\rm GeV} \\
1.2~~~ \leq & \tan \beta & \leq 60, \\
& {\rm sgn}(A^{\prime}), {\rm sgn}(\mu^{\prime}) & =-1,+1~.
\end{array}
\label{eq:scan_NH}
\end{equation}
where $\Lambda_{5,3,8}$ and $M_{{\rm Mess}}$ are the SUSY breaking parameters as defined in the previous section. $\tan\beta$ is parametrized by VEVs of the MSSM Higgs fields as $\tan\beta \equiv v_{u}/v_{d}$, in which $v_{u}$ and $v_{d}$ are VEVs of $H_{u}$ and $H_{d}$, respectively. Based on Eq.(\ref{eq:NHmess}), the magnitude of the NH terms can lie from about $10^{-30}$ GeV up to about 200 TeV with these ranges of the fundamental parameters. We also should note that we parametrize the strength of the NH trilinear interactions as $T^{\prime}_{i} \equiv A_{i}^{\prime}y_{i}$. With this setup, we keep the CKM matrix in rotation of the squark mass-square matrices, and also the NH terms do not lead to charged lepton flavour violation. Even though the numerical values of the NH terms are fixed with $\Lambda_{i}$ and $M_{{\rm Mess}}$, we assign positive and negative signs to the NH terms randomly.

The data analyzed in our work were generated by using SPheno-4.0.4 \cite{Porod:2003um,Goodsell:2014bna} numerical calculation package, which was obtained with SARAH \cite{Staub:2008uz,Staub:2015iza}. While scanning the parameter space, we use the Metropolis-Hastings algorithm \cite{Baer:2008jn,Belanger:2009ti}. We have modified the SARAH package such that  the NH terms can be imposed at the messenger scale at which the SSB terms are induced as defined in Eqs.(\ref{eq:massgaugionos}, \ref{eq:sfermionmasses}). We performed random scans over the fundamental parameters within their ranges given in Eq.(\ref{eq:scan_NH}). SPheno first runs the 1-loop RGEs for the Yukawa and gauge couplings from $M_{Z}$ to $M_{{\rm Mess}}$. Once the numerical values are determined at $M_{{\rm Mess}}$, all the SSB parameters parametrized with $\Lambda_{i}$ and $M_{{\rm Mess}}$ are imposed and SPheno evolves masses and couplings back to $M_{{\rm SUSY}}$ through the 2-loop RGEs, where $M_{{\rm SUSY}} \equiv \sqrt{m_{\tilde{t}_{L}}m_{\tilde{t}_{R}}}$ is the scale at which the supersymmetric particles decouple from the spectrum. The results from SPheno runs include the masses of the supersymmetric particles and Higgs bosons, couplings as well as the branching ratios for the on-shell decay processes. We accept only the solutions which do not yield negative mass-square for the scalar particles and consistent with the the radiative electroweak symmetry breaking (REWSB), which fixes the bilinear Higgs mixing term $\mu$ with $m_{H_{u}}$, $m_{H_{d}}$ and $\tan\beta$ to be consistent with the electroweak symmetry breaking.

After the data set is generated, we apply some experimental constraints successively from the mass bounds on the supersymmetric particles \cite{ParticleDataGroup:2014cgo,ATLAS:2021twp,ATLAS:2020syg,ATLAS:2022rcw} and the Higgs boson \cite{ATLAS:2012yve,CMS:2012qbp,CMS:2013btf}, rare decays of $B-$meson \cite{Belle-II:2022hys,CMS:2020rox}. These constraints can be summarized as follows:

\begin{equation}
\begin{array}{rcl}
&m_{\tilde{g}} &\geq 2.2~~ {\rm TeV} \\
123 \leq& m_{h} &\leq 127~~ {\rm GeV} \\
2.9\times 10^{-4} \leq & BR (b \rightarrow s {\gamma}) &\leq 3.87\times 10^{-4} ~~(2\sigma)\\
1.95\times 10^{-9} \leq &BR (B_s \rightarrow {\mu}^{+} {\mu}^{-})&\leq 3.43 \times 10^{-9} ~~~ (2\sigma)
\end{array}
\label{eq:constraints}
\end{equation}

In addition to these constraints, after the recent updates in muon $g-2$, we also require our solutions do not spoil the muon $g-2$ results with large or negative SUSY contributions by imposing the constraint as $0 \leq \Delta a_{\mu} \leq 6.6\times 10^{-10}$ \cite{Aliberti:2025beg,Muong-2:2025xyk}.

As is well known the GMSB models usually yield gravitino to be the lightest supersymmetric particle (LSP) in the spectrum. However, the models in the GMSB-ADJ class can also lead to MSSM neutralinos to be LSP in a small portion of their fundamental parameter space \cite{Gogoladze:2016grr}. These two different cases of LSP lead to different phenomenology, and hence, they might be subjected to different experimental results, when they are considered to be candidates of dark matter (DM). Considering the possible NH contributions in the DM phenomenology, we accept only the solutions which accommodate either gravitino or one of the MSSM neutralinos to be LSP in the spectrum. However, despite including in our analyses, we do not observe any significant effect from the NH contributions in the DM implications of the models; and hence, we leave it out of the scope of our work.

\section{Fundamental Parameter Space of GMSB-ADJ}
\label{sec:fund}

\begin{figure}[t!]
\begin{subfigure}{0.5\textwidth}
\centering
Holomorphic
\includegraphics[scale=0.4]{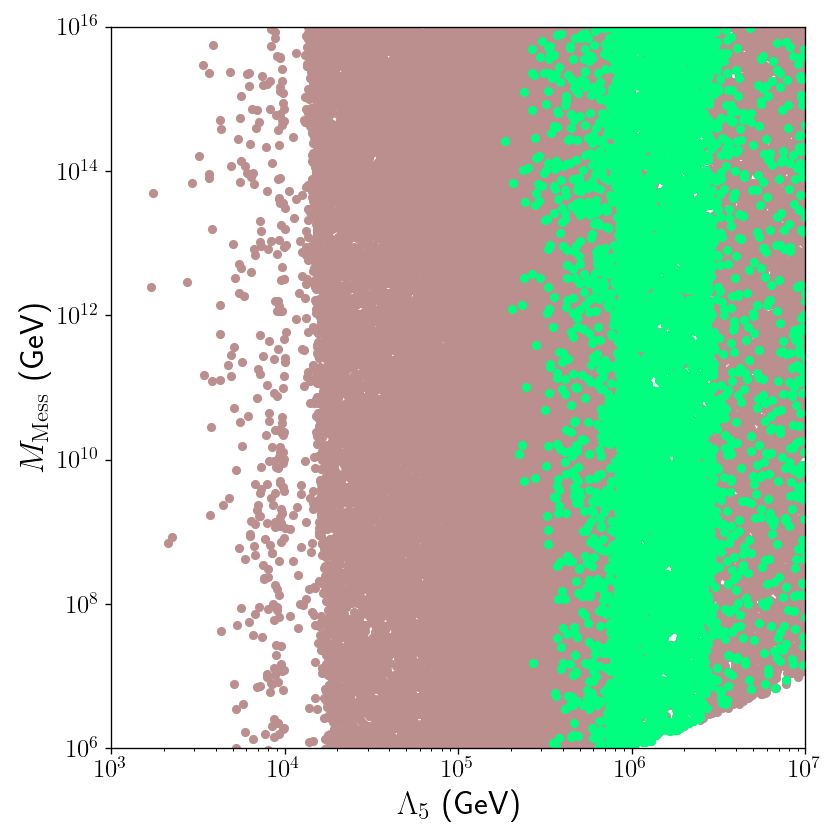}
\includegraphics[scale=0.4]{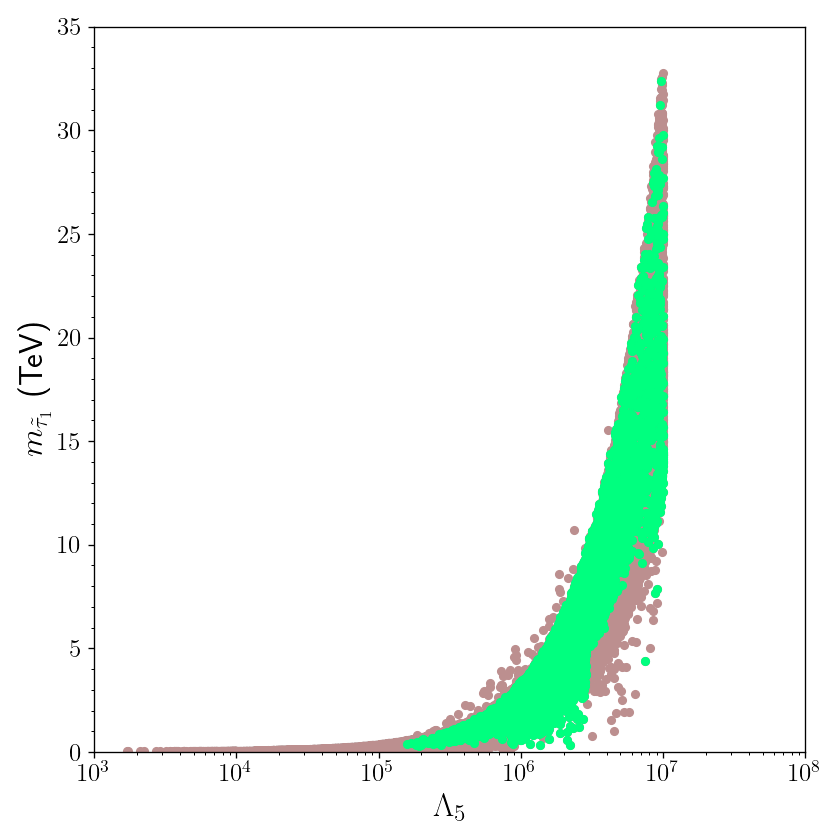}
\end{subfigure}%
\begin{subfigure}{0.5\textwidth}
\centering
NH
\includegraphics[scale=0.4]{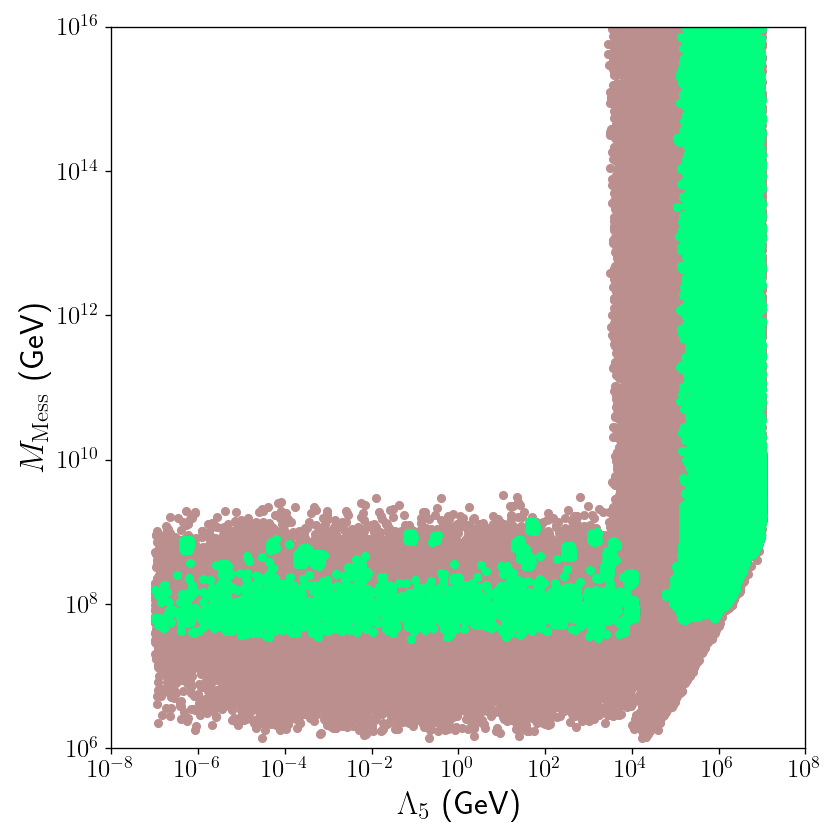}
\includegraphics[scale=0.4]{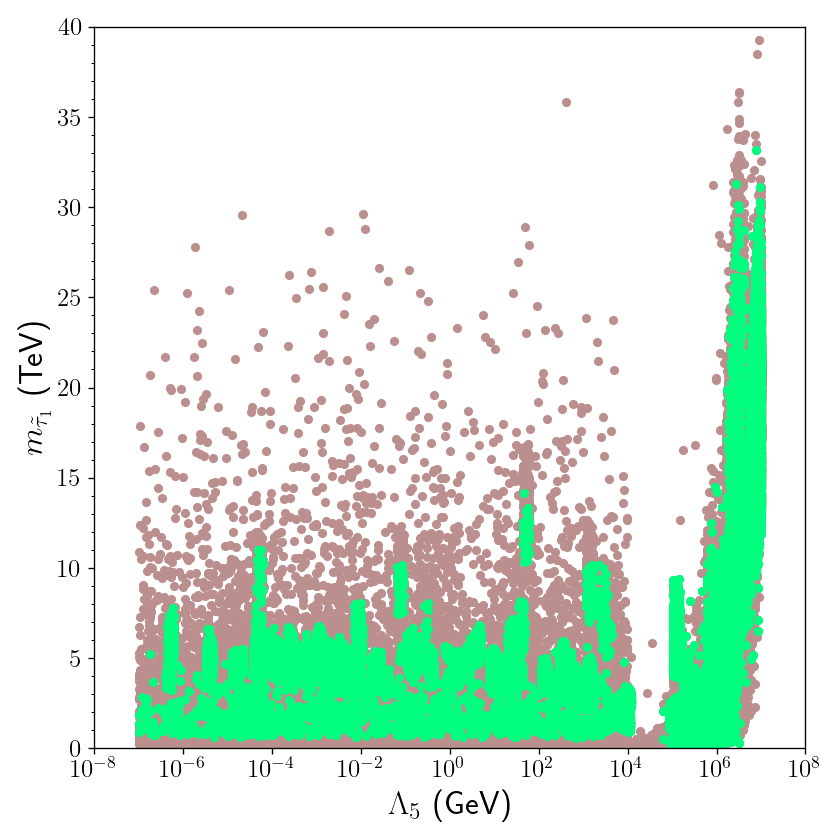}
\end{subfigure}
\caption{The allowed ranges of the fundamental parameters $M_{{\rm Mess}}$ and $\Lambda_{5}$ in the absence of NH terms (left) and with the NH contributions (right). The bottom planes display the results for the stau mass, which is crucial to shape the ranges of $\Lambda_{5}$. All the points are compatible with the REWSB and condition on LSP to be either Gravitino or Neutralino. Green points are compatible with the constraints listed in Eq.(\ref{eq:constraints}).}
\label{fig:L5MMess_staus}
\end{figure}

\begin{figure}[t!]
\includegraphics[scale=0.4]{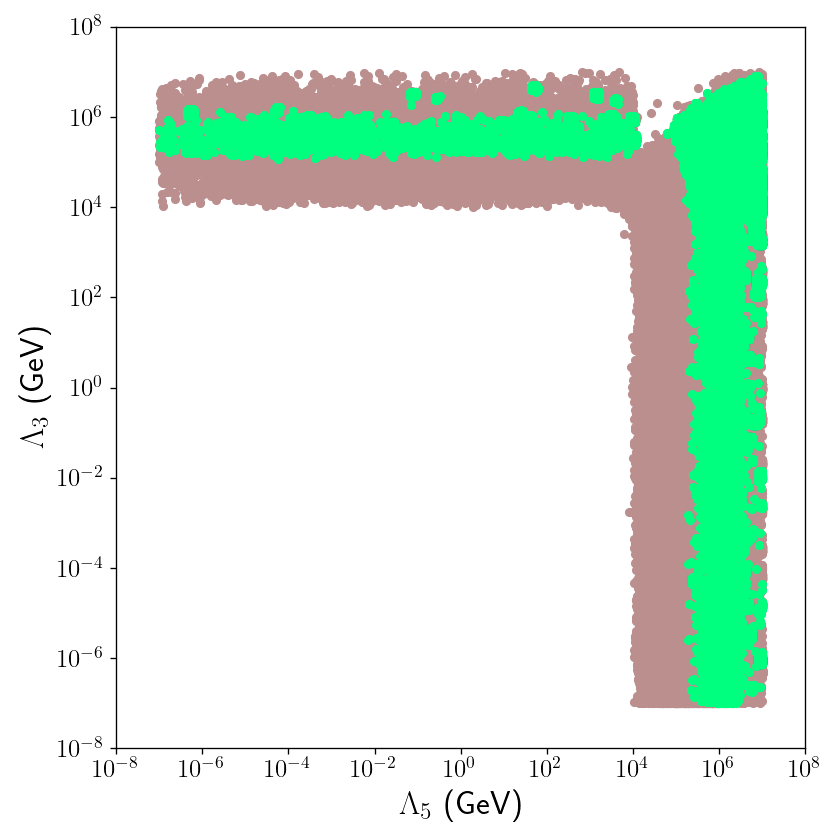}%
\includegraphics[scale=0.4]{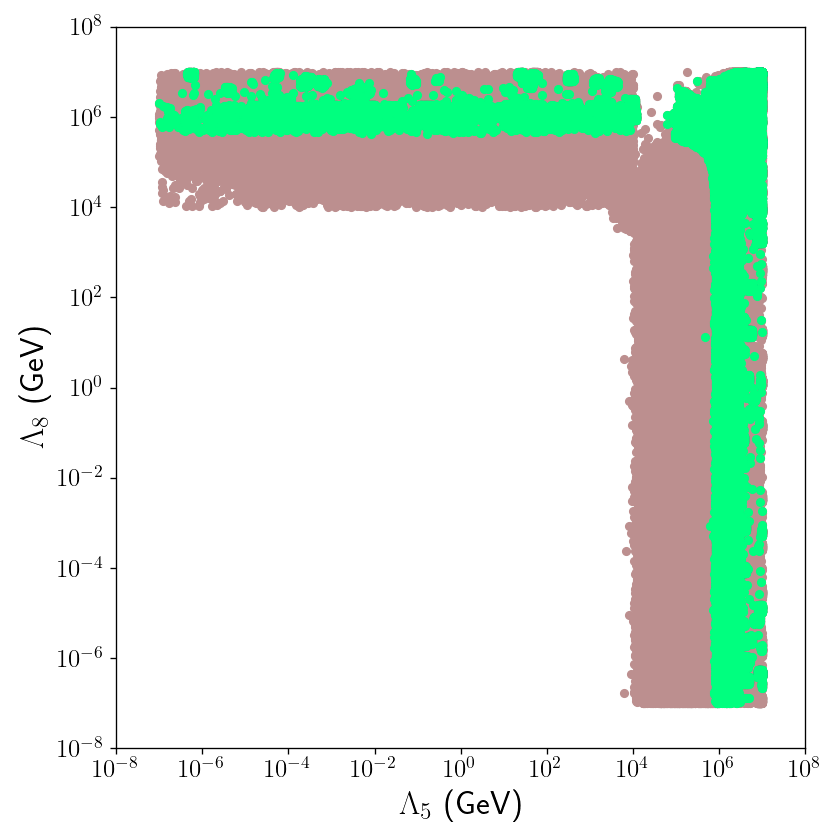}
\includegraphics[scale=0.4]{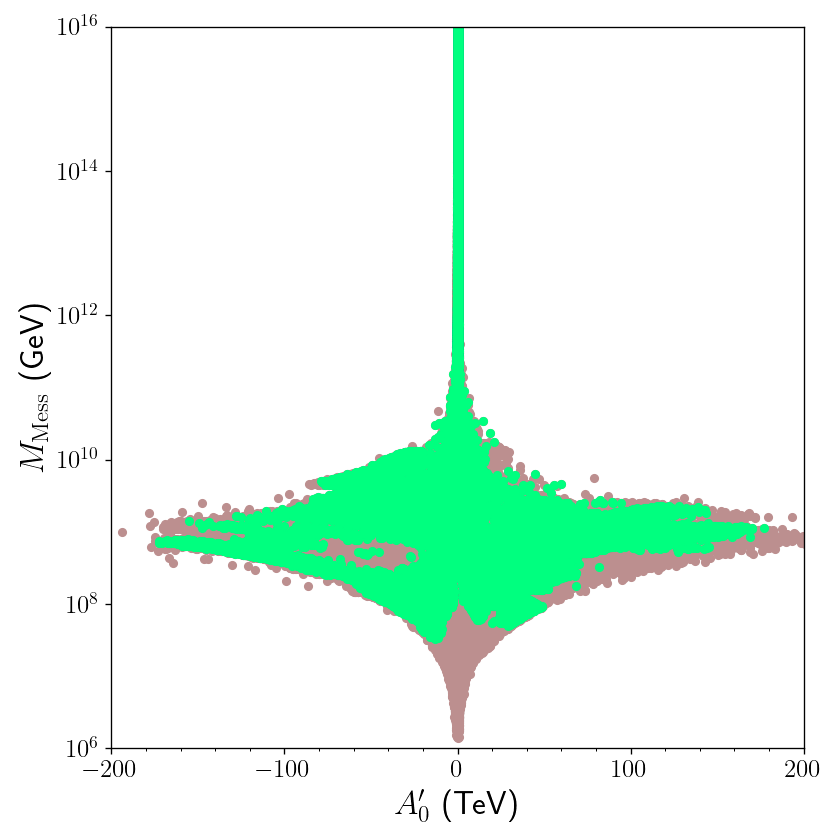}
\includegraphics[scale=0.4]{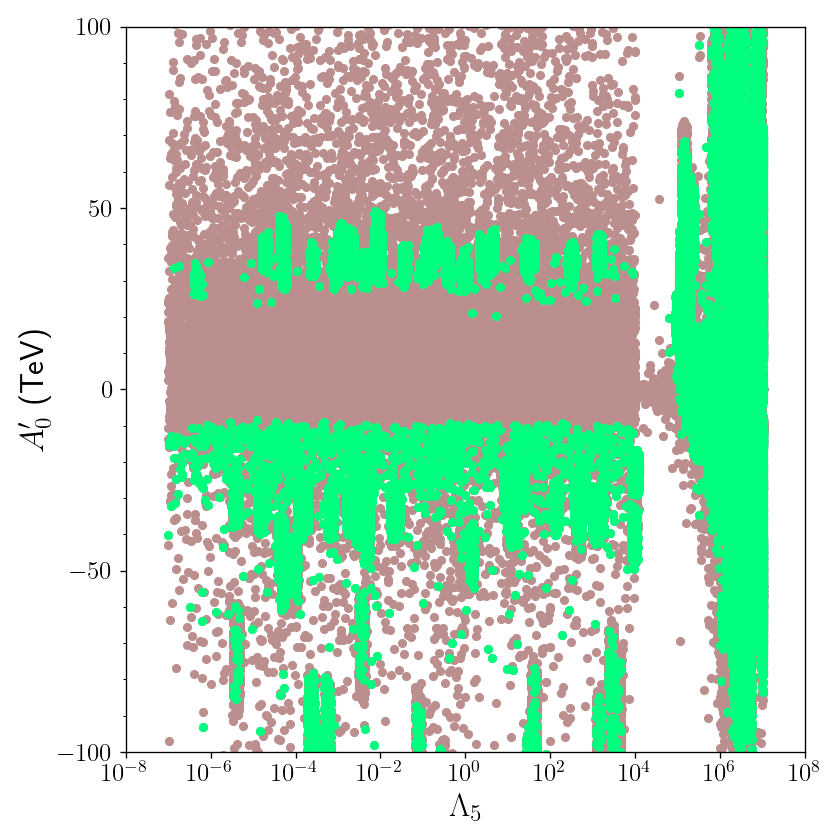}
\caption{Plots in the $\Lambda_{3}-\Lambda_{5}$, $\Lambda_{3}-\Lambda_{8}$, $M_{{\rm Mess}}-A_{0}^{\prime}$, and $\Lambda_{5}-A_{0}^{\prime}$ planes. The color coding is the same as in Figure \ref{fig:L5MMess_staus}.}
\label{fig:GMSBfunds}
\end{figure}

We first display the allowed ranges for the fundamental parameters $M_{{\rm Mess}}$ and $\Lambda_{5}$ in Figure \ref{fig:L5MMess_staus}. All the points are compatible with the REWSB and condition on LSP to be either Gravitino or Neutralino. Green points are compatible with the constraints listed in Eq.(\ref{eq:constraints}). The right planes depict the results for the case in which the NH terms are absent, while the right planes are generated for the models including the NH contributions. Note that these two data sets are generated independently. As seen from the the $M_{{\rm Mess}}-\Lambda_{5}$ plane for the holomorphic case (top-left), $\Lambda_{5}$ cannot be smaller than about $10^{3}$ GeV, and the solutions consistent with the experimental constraints employed in our analyses can be realized when $\Lambda_{5}\gtrsim 2\times 10^{5}$ GeV, while $M_{{\rm Mess}}$ can lie in a wide range from $10^{6}$ to $10^{16}$ GeV. This bound on $\Lambda_{5}$ arises from the stau mass. As discussed before, the right-handed stau mass is induced only through the interactions with $\Sigma_{5}$, and its SSB mass depends only on $\Lambda_{5}$. Therefore, small ranges of $\Lambda_{5}$ yield tachyonic states for the right-handed stau when $\Lambda_{5} \lesssim 10^{3}$ GeV. Besides, despite avoiding the tachyonic states, stau happens inconsistently light for $10^{3} \lesssim \Lambda_{5} \lesssim 10^{5}$ GeV. These results can be seen in the $m_{\tilde{\tau}_{1}}-\Lambda_{5}$ plane for the holomorphic case (bottom-left). 

The right planes show the results in the same planes for the solutions in which the NH contributions are also taken into account. The impact from such contributions can be directly seen in the regions with small $\Lambda_{5}$. As seen from the top-right $M_{{\rm Mess}}-\Lambda_{5}$, one can find solutions even in the regions with $\Lambda_{5} \gtrsim 10^{-7}$ GeV. Clearly, the NH contributions in this region are large enough that the right-handed stau mass-square remains positive. However, the experimental mass bound on stau restricts $M_{{\rm Mess}}$ to be larger than about $10^{8}$ GeV. We also observe an upper bound on $M_{{\rm Mess}}$ in this region as $M_{{\rm Mess}} \lesssim 10^{9}$ GeV. Above this region there is no solution at all, because the right-handed stau mass turns to be tachyonic again. Indeed, larger $M_{{\rm Mess}}$ means also RGE evolution in a larger energy interval. In this region, the Bino mass ($M_{1}$) also turns to be negative, and it starts reducing the stau mass through RGEs. A longer RG evolution lead $M_{1}$ to enhance its decreasing effect on the stau mass. In this case, although stau mass-square starts from positive, it falls into the negative values through its RG evolution when $M_{{\rm Mess}} \gtrsim 10^{9}$ GeV, and hence, it puts an upper bound on $M_{{\rm Mess}}$. We also display the stau mass in the parameter space in the bottom-right plane, which shows that the stau can be as heavy as about 15 TeV in this region.

The results shown in Figure \ref{fig:L5MMess_staus} in comparison with the holomorphic case reveal the significant impact from the NH terms. In the small $\Lambda_{5}$ region, following Eq.(\ref{eq:NHmess}), at least one of the messenger fields should be heavy to induce relatively large NH terms. This requirement can be seen from the planes of Figure \ref{fig:GMSBfunds}, whose color coding is the same as in Figure \ref{fig:L5MMess_staus}. The top planes display the ranges of $\Lambda_{3}$ (left) and $\Lambda_{8}$ (right) in comparison with $\Lambda_{5}$. As seen from the results in these planes, both $\Lambda_{3}$ and $\Lambda_{8}$ should be larger than about $10^{4}$ in the regions with small $\Lambda_{5}$ to avoid tachyonic stau states. The consistent solutions (green) shift the scales for these parameters further to about $10^{5}$ GeV. The order of the NH terms also inversely depends on $M_{{\rm Mess}}$, which requires relatively small values for the SUSY breaking scale. The $M_{{\rm Mess}}-A_{0}^{\prime}$ plane shows that $M_{{\rm Mess}}$ cannot be larger than about $10^{11}$ GeV in order to induce sizeable NH terms. The solutions in this plane shows that $A_{0}^{\prime}$ can be as large as about 200 TeV. However, these solutions cannot provide consistently heavy staus when $-10 \lesssim A_{0}^{\prime} \lesssim 20$ TeV, as seen from the $\Lambda_{5}-A_{0}^{\prime}$ plane. Indeed, heavy stau masses mostly require negative $A_{0}^{\prime}$, and heavy staus ($\sim 15$ TeV) can be realized when $A_{0}^{\prime} \sim -200$ TeV.

\subsection{Supersymmetric Mass Spectra}
\label{subsec:susymass}

\begin{figure}[t!]
\includegraphics[scale=0.4]{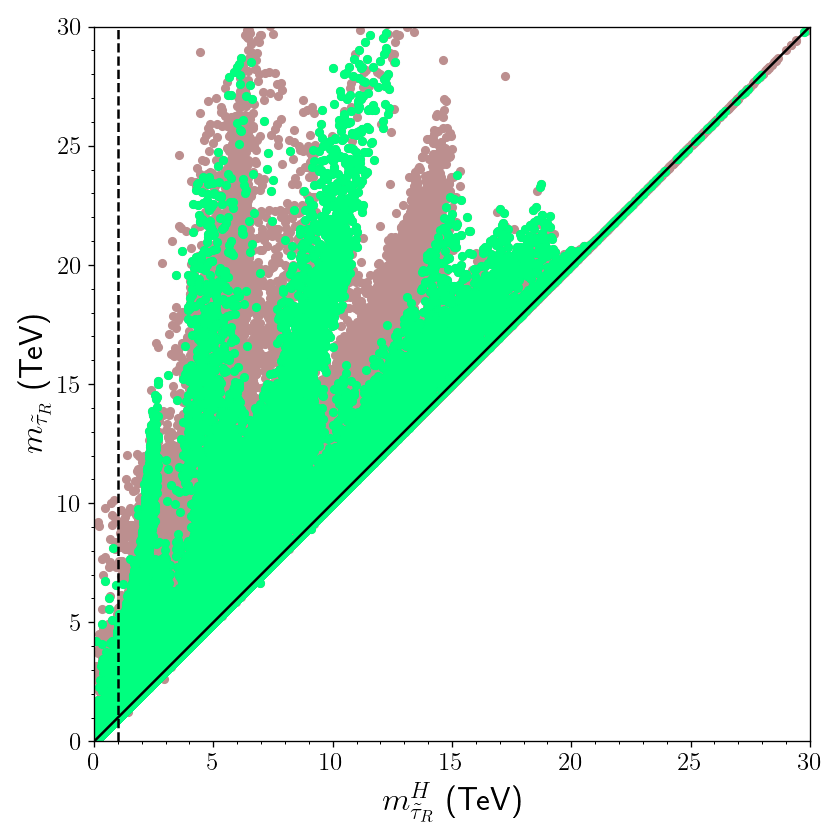}%
\includegraphics[scale=0.4]{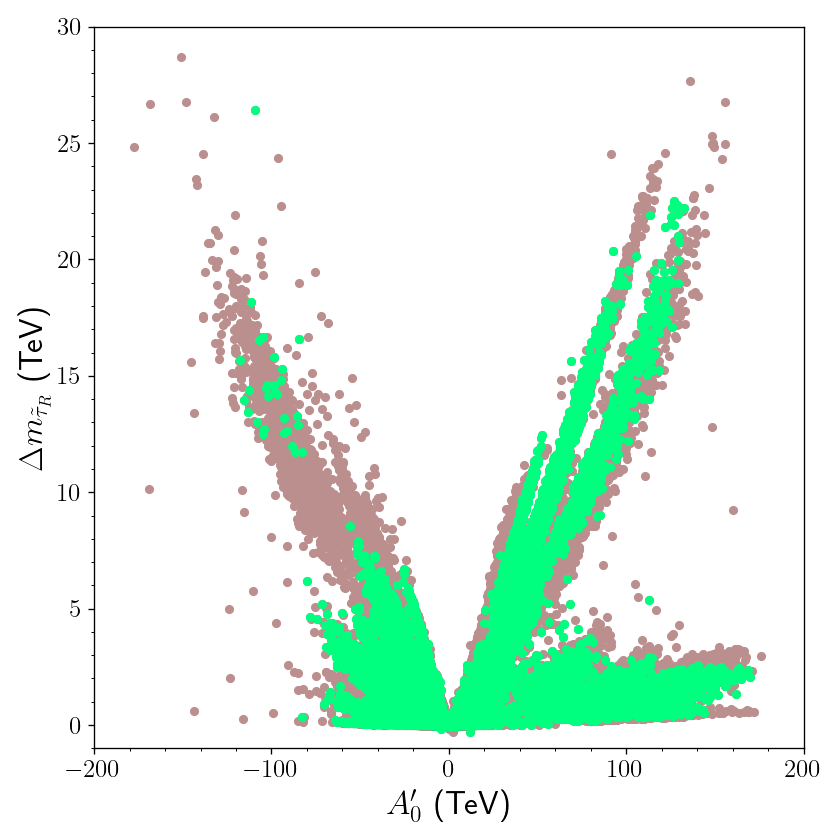}
\caption{The right-handed stau masses in holomorphic ($m_{\tilde{\tau}_{R}}^{H}$) and NH cases (left), the difference in its mass due to the NH contributions in correlation with its mixing (right). The color coding is the same as in Figure \ref{fig:GMSBfunds}. The diagonal line in the left plane depicts the solutions for negligible or zero NH contributions to the stau mass ($m_{\tilde{\tau}_{R}}=m_{\tilde{\tau}_{R}}^{H}$). The dashed vertical line is also used to indicate the solutions which lead to light staus in the holomorphic case ($m_{\tilde{\tau}_{R}} \leq 1$ TeV)}
\label{fig:GMSBD_staus}
\end{figure}

\begin{figure}[t!]
\includegraphics[scale=0.4]{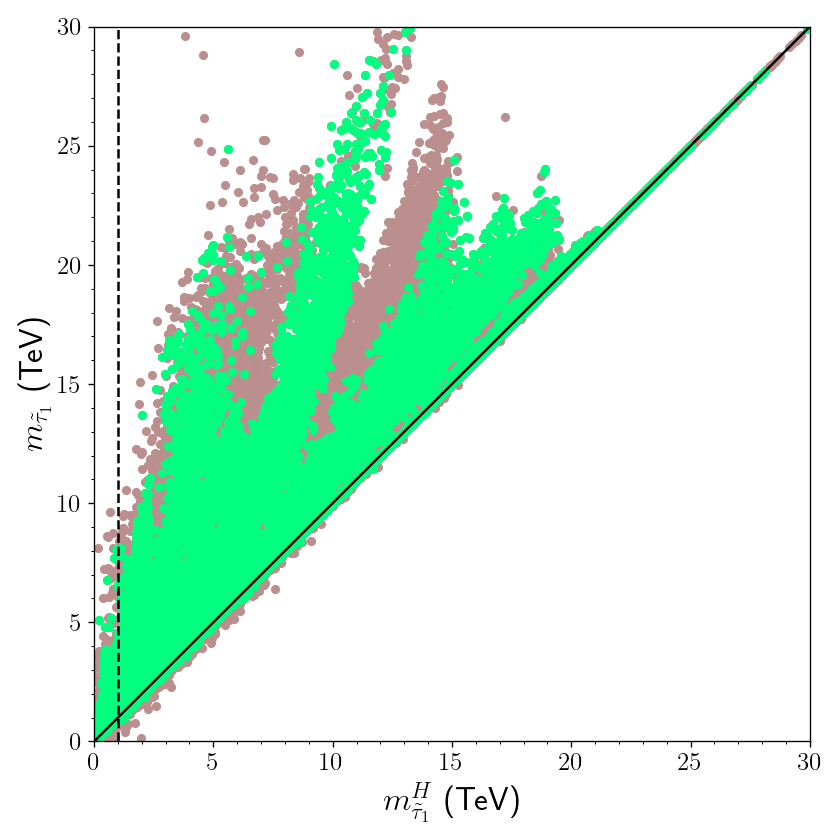}%
\includegraphics[scale=0.4]{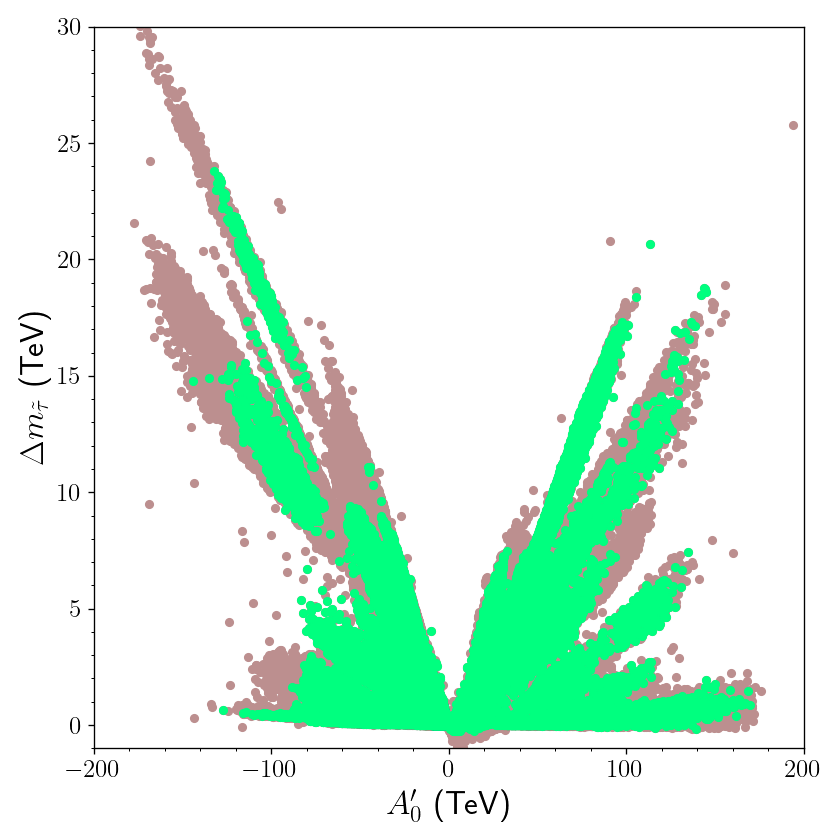}
\caption{Plots in the $m_{\tilde{\tau}_{1}} - m_{\tilde{\tau}_{1}}^{H}$ and $\Delta m_{\tilde{\tau}_{1}}-A_{0}^{\prime}$ planes. The color coding is the same as in Figure \ref{fig:L5MMess_staus}. The diagonal lines in each plane depict the solutions in which the NH contributions are negligible.}
\label{fig:GMSBD_massstaus}
\end{figure}

\begin{figure}[t!]
\includegraphics[scale=0.4]{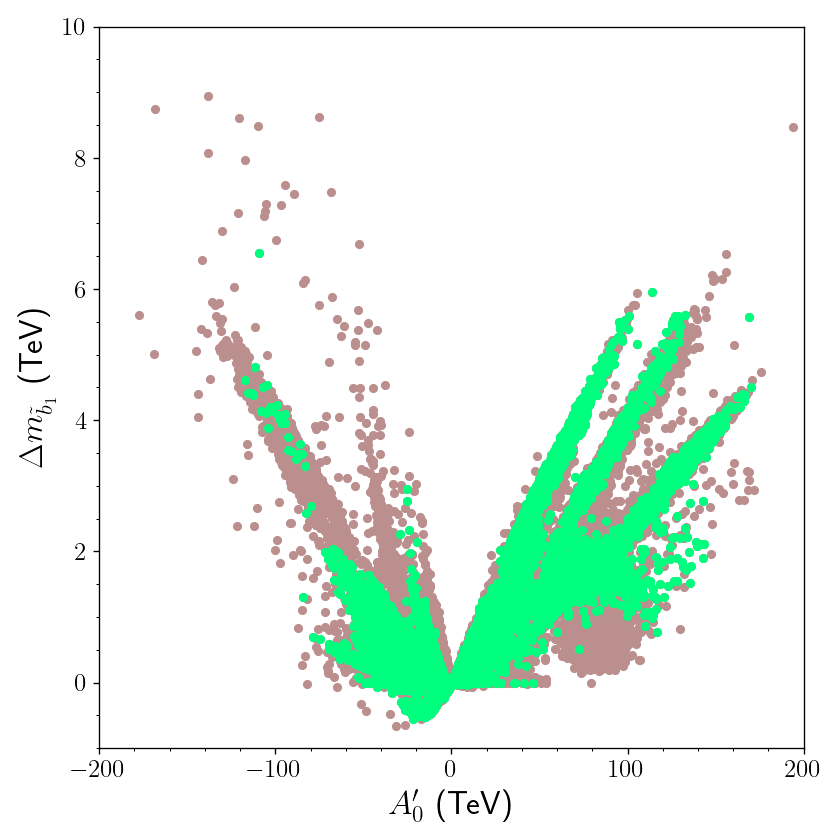}%
\includegraphics[scale=0.4]{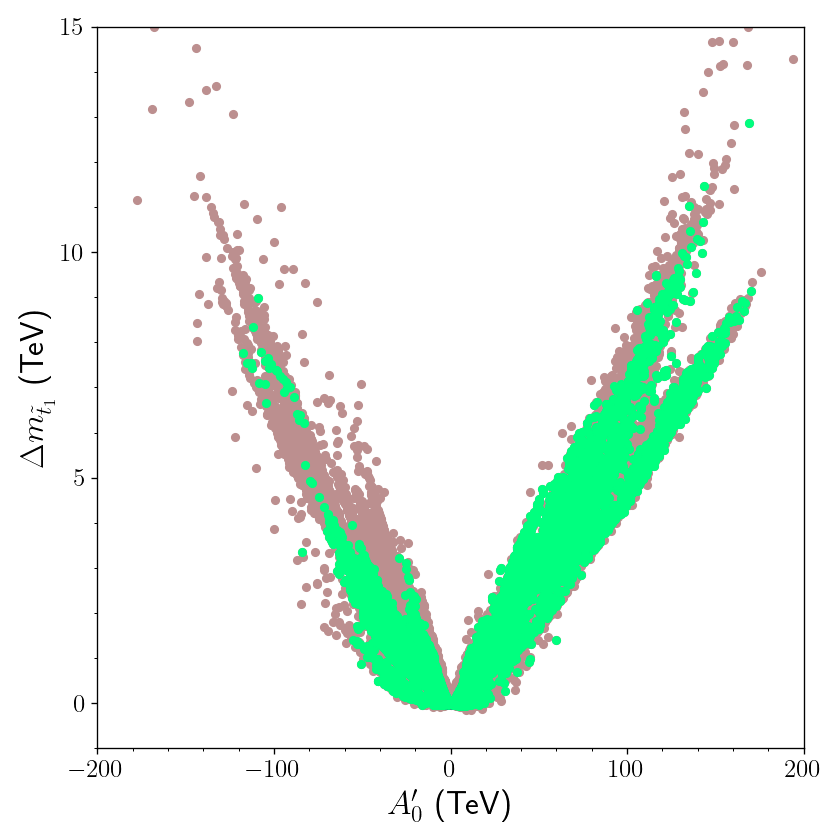}
\includegraphics[scale=0.4]{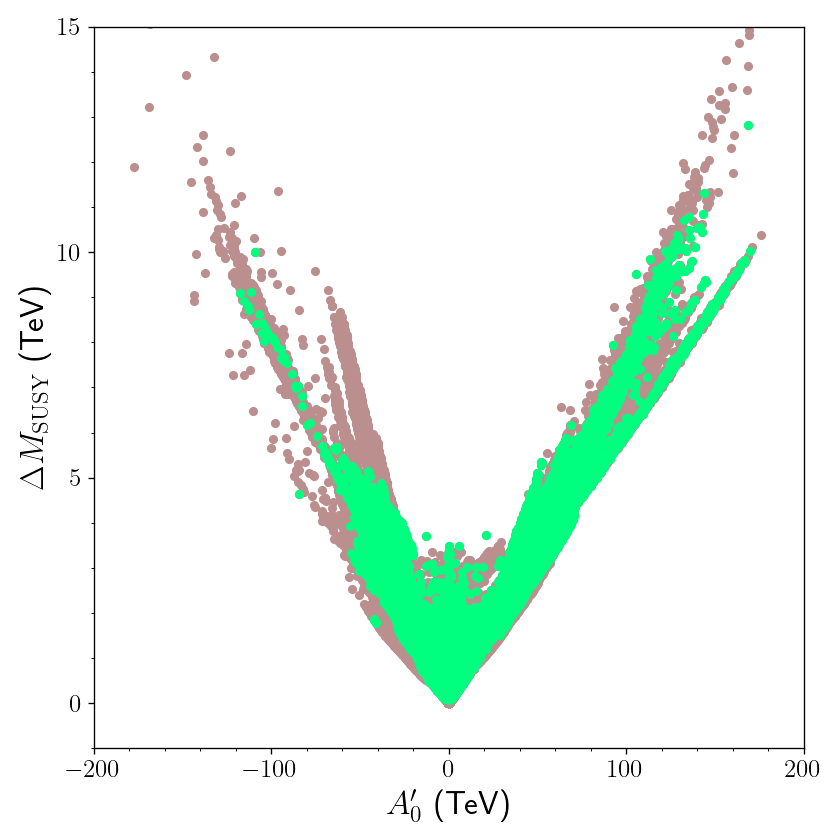}%
\includegraphics[scale=0.4]{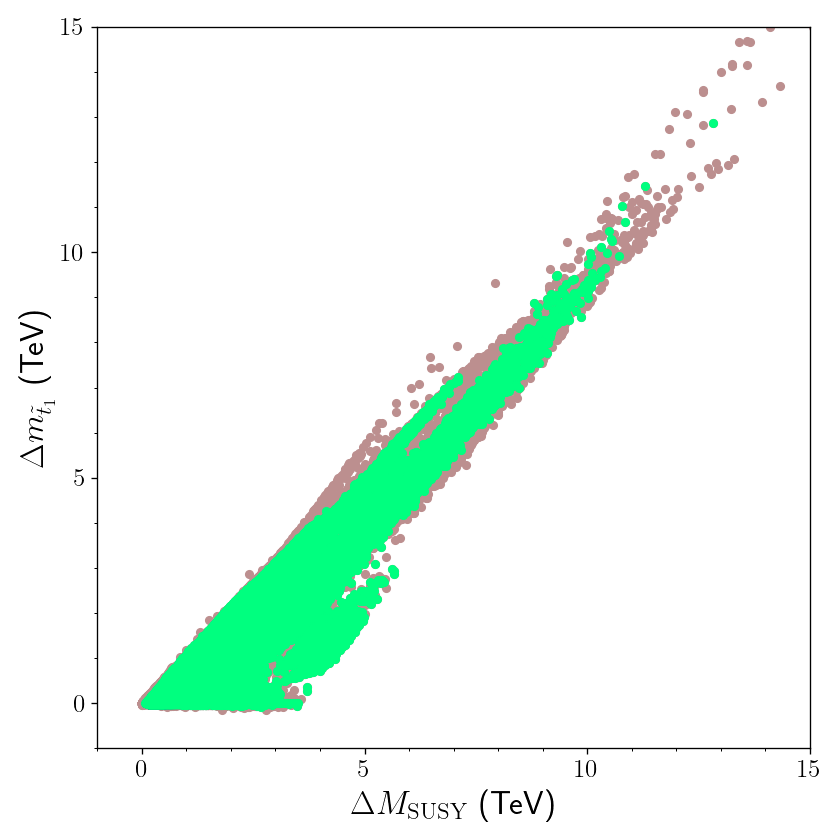}
\caption{The NH contributions to the squarks in the $\Delta m_{\tilde{b}_{1}}-A_{0}^{\prime}$, $\Delta m_{\tilde{t}_{1}}-A_{0}^{\prime}$, $\Delta M_{{\rm SUSY}}-A_{0}^{\prime}$, and $\Delta m_{\tilde{t}_{1}} - \Delta M_{{\rm SUSY}}$ planes. The color coding is the same as in Figure \ref{fig:L5MMess_staus}.}
\label{fig:GMSBD_squarks}
\end{figure}

In this section, we discuss the NH contributions to the SUSY spectra by comparing the masses of SUSY particles with and without the NH terms. The solutions represented in this section are calculated twice one of which employs the NH terms, while the other set them to zero for the same input values (we call it the holomorphic case). Note that these results depict the solutions which do not lead to tachyonic states when the NH contributions are switched off, therefore the regions with $\Lambda_{5} \lesssim 10^{3}$ GeV are not represented in these planes. The parameters calculated without the NH contributions are denoted with the upper index ``H". 

We discussed the crucial role of the stau mass in the previous section, so it might be better to show first the NH contributions to the right-handed stau mass. Figure \ref{fig:GMSBD_staus} displays our results for the comparison in stau mass between the holomorphic and NH cases in the $m_{\tilde{\tau}_{R}}-m_{\tilde{\tau}_{R}}^{H}$ and $\Delta m_{\tilde{\tau}_{R}}-A_{0}^{\prime}$, where $\Delta m_{\tilde{\tau}_{R}} = m_{\tilde{\tau}_{R}}-m_{\tilde{\tau}_{R}}^{H}$. The color coding is the same as in Figure \ref{fig:GMSBfunds}. The diagonal line in the left plane depicts the solutions for negligible or zero NH contributions to the stau mass ($m_{\tilde{\tau}_{R}}=m_{\tilde{\tau}_{R}}^{H}$). As is seen, the light stau solutions (say $\lesssim 1$ TeV) correspond to small $\Lambda_{5}$ ($\lesssim 10^{5}$ GeV) in the holomorphic case can be as heavy as about 10 TeV in this region with $m_{\tilde{\tau}_{R}}^{H} \lesssim 1$ TeV. Besides, the stau can be as heavy as about 30 TeV with the NH contributions, while $m_{\tilde{\tau}_{R}}^{H} \lesssim 20$ TeV. The NH contributions become negligible in the regions when the holomorphic case involves staus heavier than about 20 TeV in the spectra, as observed that all the solutions are accumulated around the diagonal line when $m_{\tilde{\tau}_{R}}^{H} \gtrsim 20$ TeV. The correlation in the NH contributions to the stau mass with the NH terms is shown in the $\Delta m_{\tilde{\tau}_{R}}-A_{0}^{\prime}$. As expected, the large $A_{0}^{\prime}$ values contribute most such that one can realize about 25 TeV difference in the stau mass when $|A_{0}^{\prime}| \gtrsim 100$ TeV. Even though it seems the NH terms always contribute to the stau mass positively, there is a small region where they reduce the stau mass. However the difference in these solutions does not exceed 1 TeV so that it is not very visible in Figure \ref{fig:GMSBD_staus}.

Note that the right-handed staus receive the NH contributions only through RGEs. On the other hand, the mass eigenstates also receive these contributions through the mixing in their mass-square matrix as given in Eq.(\ref{sfermionsmass2}). Figure \ref{fig:GMSBD_massstaus} represents the masses for the lightest stau mass eigenstate in comparison with the holomorphic case in the $m_{\tilde{\tau}_{1}} - m_{\tilde{\tau}_{1}}^{H}$, and the NH contributions in the $\Delta m_{\tilde{\tau}_{1}}-A_{0}^{\prime}$ planes. The color coding is the same as in Figure \ref{fig:L5MMess_staus}. The diagonal lines in the left plane depicts the solutions in which the NH contributions are negligible. The solutions in the $m_{\tilde{\tau}_{1}} - m_{\tilde{\tau}_{1}}^{H}$ shows a similar behaviour as mass of the flavour eigenstate $m_{\tilde{\tau}_{R}}$. On the other hand, we observe that the difference from the holomorphic case is slightly smaller for the mass eigenstates. The heaviest stau mass can be realized at about 30 TeV, which corresponds to about 20 TeV mass difference from the holomorphic case at most. The solutions which yield to light staus ($m_{\tilde{\tau}_{1}}^{H}\lesssim 1$ TeV) in the holomorphic case, can accommodate staus as heavy as about 7-8 TeV with the NH contributions. The correlation with the NH terms is seen to be quite similar for $\Delta m_{\tilde{\tau}_{1}}$ in the $\Delta m_{\tilde{\tau}_{1}}-A_{0}^{\prime}$ plane.

We conclude the NH contributions to the SUSY spectra by investigating their effects on the squark masses in Figure \ref{fig:GMSBD_squarks} with the $\Delta m_{\tilde{b}_{1}}-A_{0}^{\prime}$, $\Delta m_{\tilde{t}_{1}}-A_{0}^{\prime}$, $\Delta M_{{\rm SUSY}}-A_{0}^{\prime}$, and $\Delta m_{\tilde{t}_{1}} - \Delta M_{{\rm SUSY}}$ planes. The color coding is the same as in Figure \ref{fig:L5MMess_staus}. Even though one expects a similar effect on the sbottom mass, the $\Delta m_{\tilde{b}_{1}}-A_{0}^{\prime}$ plane reveals smaller changes in the sbottom mass due to the NH contributions. It is because the large $\Lambda_{8}$ values, which is required by the heavy mass bounds on the squarks and gluinos from the experimental analyses. As we concluded from results in the $m_{\tilde{\tau}_{R}}-m_{\tilde{\tau}_{R}}^{H}$ plane that the NH contributions become negligible in the heavy mass spectra. Large $\Lambda_{8}$ values lead to heavy sbottoms and the NH contributions are as large as only about 7 TeV. On the other hand, an interesting results can be observed in the $\Delta m_{\tilde{t}_{1}}-A_{0}^{\prime}$ plane that the stop mass can differ by about 13-14 TeV. The NH effects through mixing in stop sector are suppressed by $\tan\beta$, so we do not observe a strong reducing effect from the NH contributions in their mixing. On the other hand, the difference in the stop mass modifies the renormalization scale in a way that heavy stops lead to SUSY particles decoupling from the spectra at higher energy scales, which is parametrized with $M_{{\rm SUSY}}$. A similar difference by 13-14 TeV is also realized in $M_{{\rm SUSY}}$ as shown in $\Delta M_{{\rm SUSY}}-A_{0}^{\prime}$ plane. In this case, the SUSY particles decouple from the spectra at their heavier mass scales. The modification in the renormalization scale indirectly affects all the mass spectrum. The $\Delta m_{\tilde{t}_{1}} - \Delta M_{{\rm SUSY}}$ plane reveals a nearly linear correlation between $\Delta m_{\tilde{t}_{1}}$ and $\Delta M_{{\rm SUSY}}$.

\subsection{NH Contributions to the Higgs Boson Mass}
\label{subsec:NH_Higgs}

\begin{figure}[t!]
\centering
\includegraphics[scale=0.4]{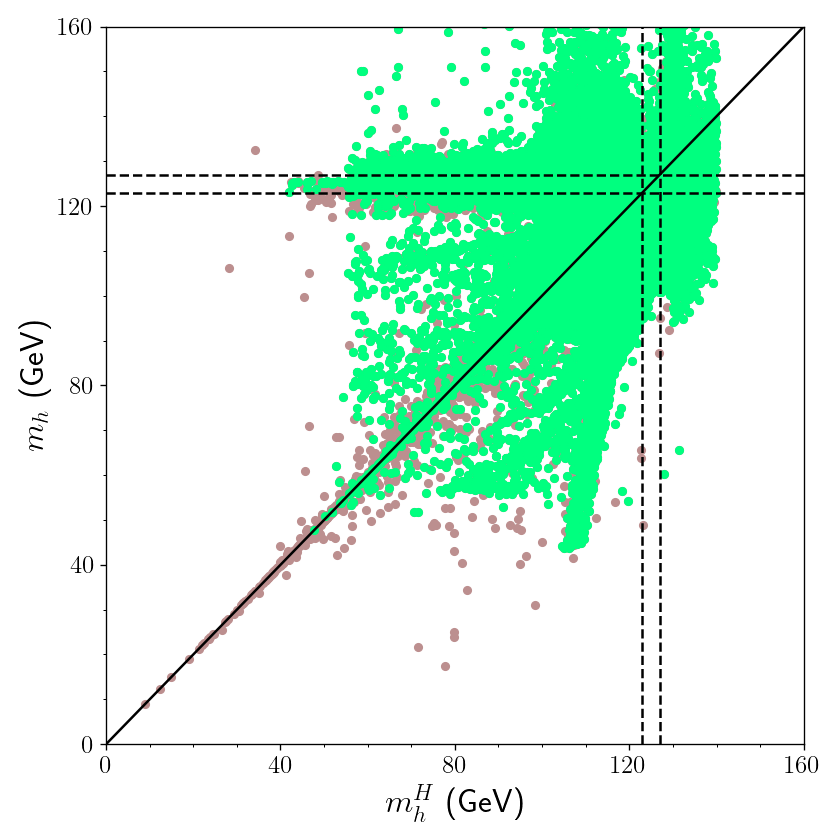}%
\includegraphics[scale=0.4]{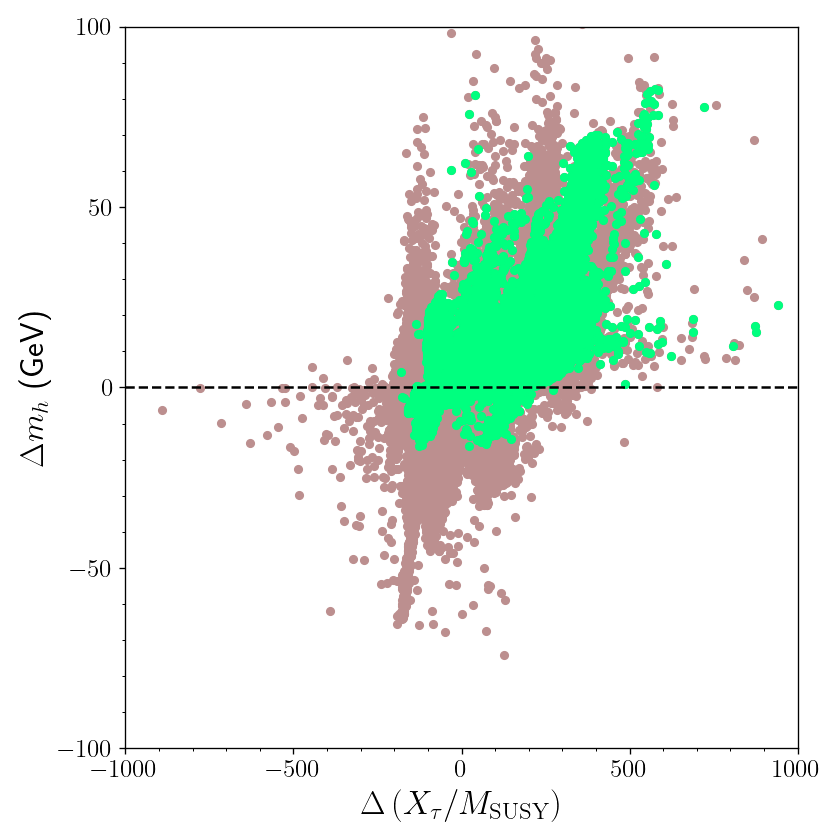}
\includegraphics[scale=0.4]{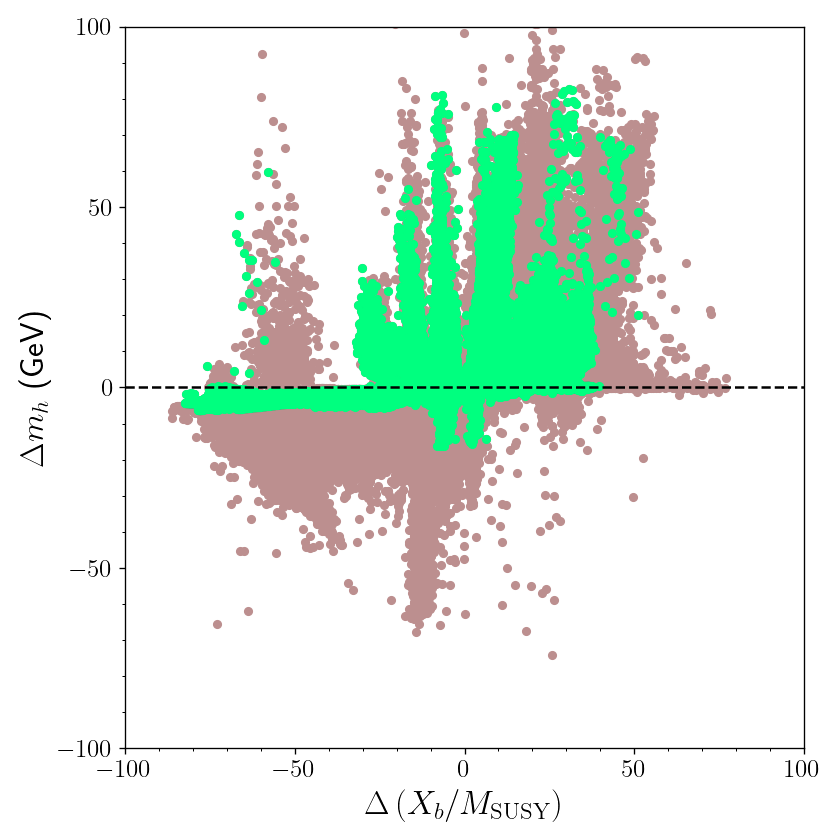}%
\includegraphics[scale=0.4]{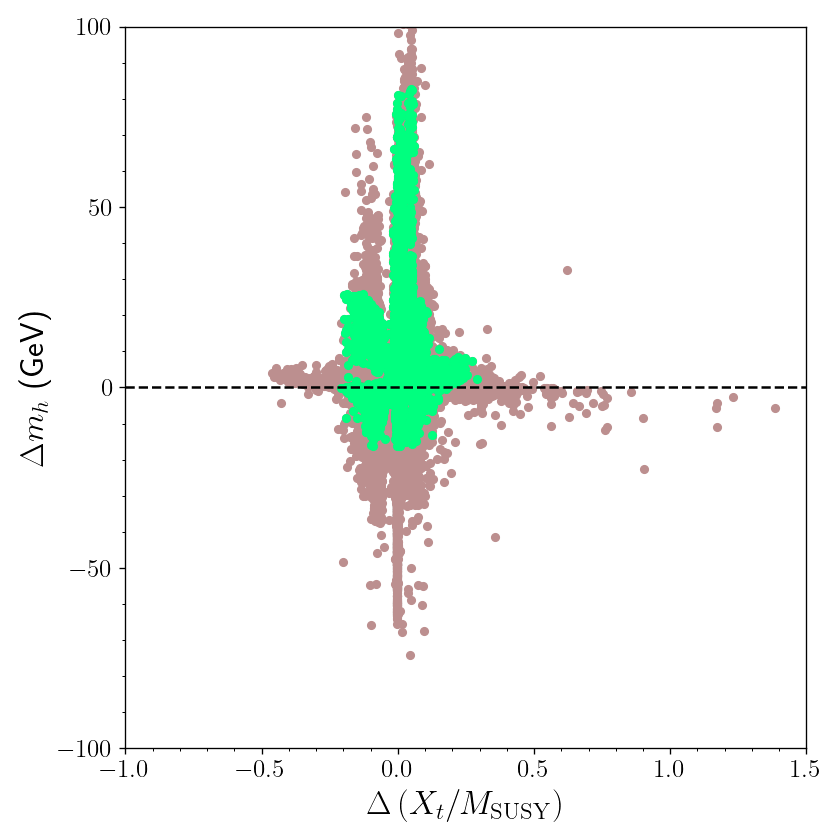}
\caption{The Higgs boson mass in comparison with the holomorphic case and the correlation between the NH contributions and mixing in sfermions. The color coding is the same as in Figure \ref{fig:L5MMess_staus}. The experimental bounds on the Higgs boson mass is not applied in the $m_{h}-m_{h}^{H}$ plane, but these bounds are rather represented by the horizontal and vertical dashed lines in this plane. The diagonal line depicts the solutions with negligible NH contributions. The solutions around the diagonal line are shown with the horizontal dashed line in the planes which plot $\Delta m_{h} \equiv m_{h}-m_{h}^{H}$.}
\label{fig:NH_Higgs}
\end{figure}

In the previous section, we observed mostly positive contributions from the NH terms to sfermions, though negative contributions can also be realized in a small region. In contrast to the positive contributions, however, the negative contributions do not exceed about 1 TeV. On the other hand, the SM-like Higgs boson mass might exhibit a significant sensitivity to the positive and negative NH contributions. We present our results for the SM-like Higgs boson in Figure \ref{fig:NH_Higgs} in comparison with the holomorphic case and the correlation between the NH contributions and mixing in sfermions. The color coding is the same as in Figure \ref{fig:L5MMess_staus}. The experimental bounds on the Higgs boson mass is not applied in the $m_{h}-m_{h}^{H}$ plane, but these bounds are rather represented by the horizontal and vertical dashed lines in this plane. The diagonal line depicts the solutions with negligible NH contributions. The solutions around the diagonal line are shown with the horizontal dashed line in the planes which plot $\Delta m_{h} \equiv m_{h}-m_{h}^{H}$. The overall mass difference in the SM-like Higgs boson can be realized as large as about 80 GeV in both cases of negative and positive NH contributions. The $m_{h}-m_{h}^{H}$ plane shows that the especially the large positive contributions can bring the SM-like Higgs boson mass to the experimentally consistent scales (green inbetween two horizontal dashed lines), even if it weighs about 40 GeV in the holomorphic case. However, the large negative contributions can happen when the both NH and holomorphic cases yield inconsistently light SM-like Higgs boson. The results in the $\Delta m_{h}-\Delta (X_{\tau}/M_{{\rm SUSY}})$ plane reveal that the large positive contributions arise from the NH contributions to the ratio $X_{\tau}/M_{{\rm SUSY}}$ such that large values for this ratio lead to large positive contributions to the Higgs boson mass, which measures about 80 GeV. On the other hand, when the contributions to this ratio are small and/or negative, the NH contributions lower the SM-like Higgs boson mass compared to the results in the holomorphic case. In the experimentally consistent region (green), the largest negative contribution to the SM-like Higgs boson mass is observed at about 20 GeV. Although it is not as large as those from the stau mixing, the sbottom mixing also minorly contributes to the Higgs boson mass, and when the stau mixing is small and negative, the NH contributions in $X_{b}/M_{{\rm SUSY}}$ slightly enhance the negative contributions to the Higgs boson mass. These solutions happen in the region with $X_{t}/M_{{\rm SUSY}} \sim 0$; however it can help driving positive contributions to the Higgs boson mass when $\Delta (X_{t}/M_{{\rm SUSY}}) \sim -0.3$, as seen from the $\Delta m_{h}-\Delta (X_{t}/M_{{\rm SUSY}})$.

\section{Notes on Muon Anomalous Magnetic Moment}
\label{sec:muong2}

\begin{figure}[t]
\centering
\includegraphics[scale=0.4]{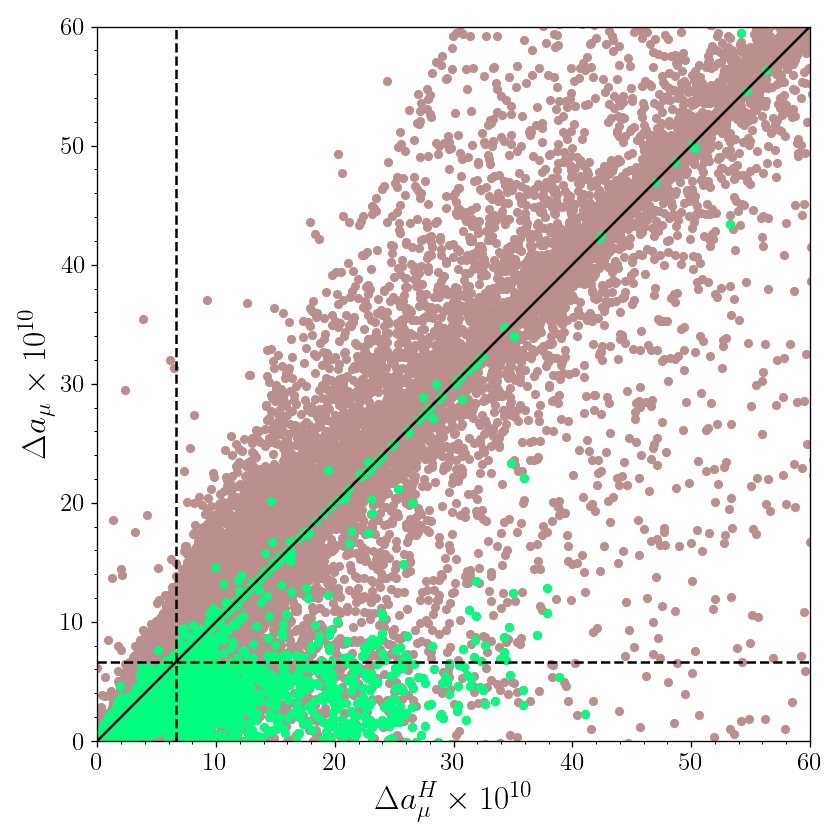}%
\includegraphics[scale=0.4]{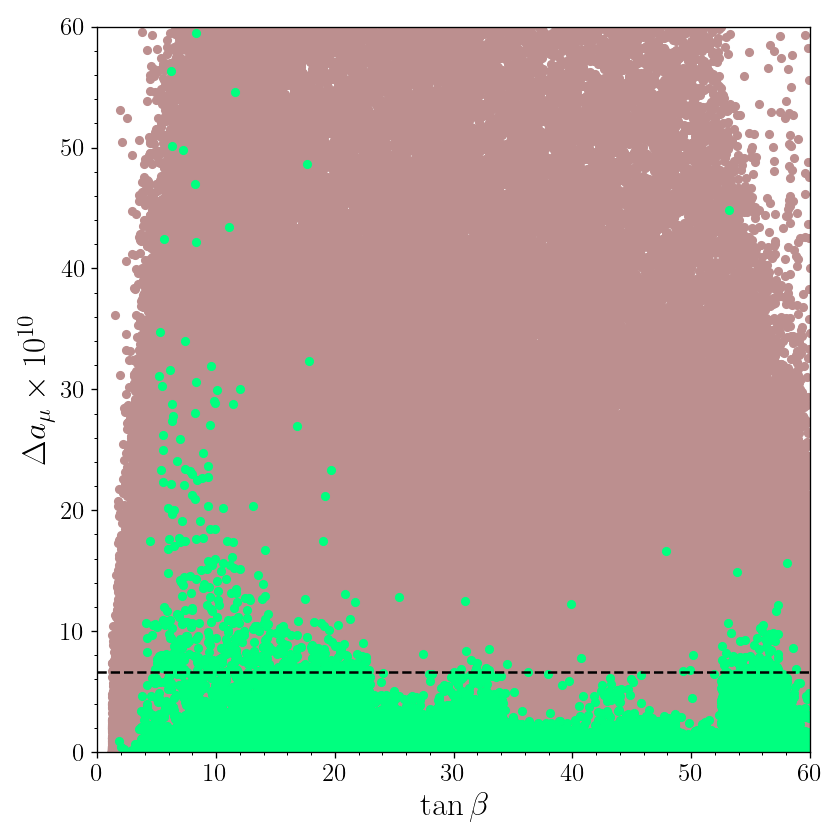}
\includegraphics[scale=0.4]{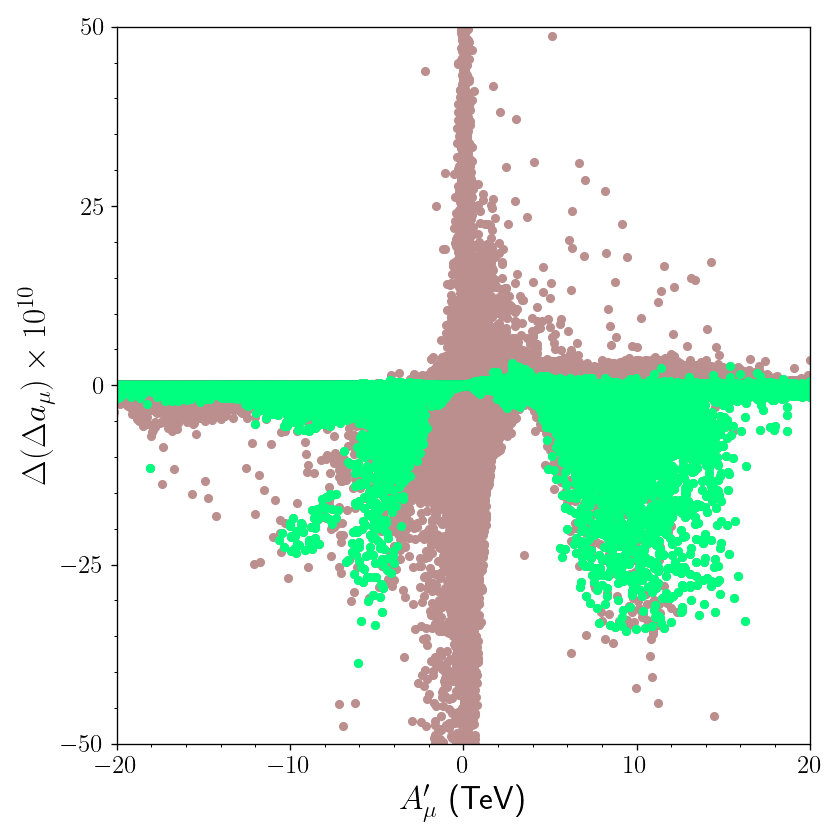}%
\includegraphics[scale=0.4]{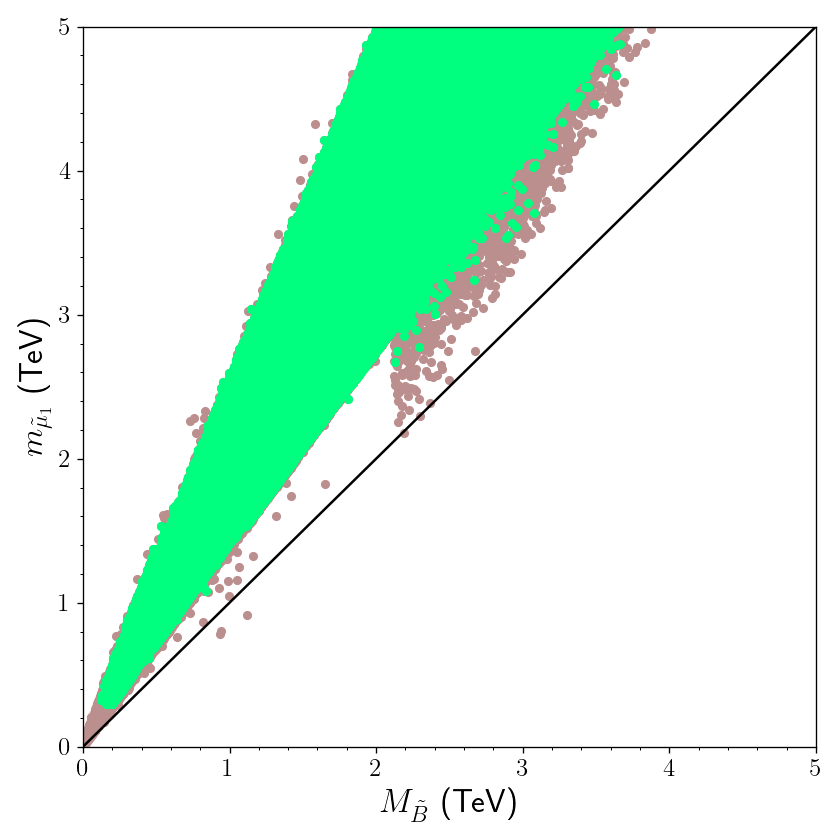}
\caption{The results for the muon anomalous magnetic moment and the NH contributions to it in the $\Delta a_{\mu}-\Delta a_{\mu}^{H}$, $\Delta a_{\mu}-\tan\beta$, $\Delta(\Delta a_{\mu})-A_{\mu}^{\prime}$, and $m_{\tilde{\mu}_{1}}-M_{\tilde{B}}$ planes. The color coding is the same as Figure \ref{fig:L5MMess_staus}. The horizontal and vertical dashed lines indicate the bound $\Delta a_{\mu} \leq 6.6\times 10^{10}$ as recently reported by the White Paper (WP25) \cite{Aliberti:2025beg}. The horizontal line in the $\Delta a_{\mu}-\Delta a_{\mu}^{H}$ plane represents the solutions with negligible NH contributions to $\Delta a_{\mu}$ ($\Delta a_{\mu} = \Delta a_{\mu}^{H}$). Also, the diagonal line in the $m_{\tilde{\mu}_{1}}-M_{\tilde{B}}$ plane indicates the mass degeneracy between smuon and Bino ($m_{\tilde{\mu}_{1}} = M_{\tilde{B}}$).}
\label{fig:NH_muong2}
\end{figure}

One of the main motivations behind this study was to analyse the NH contributions to muon $g-2$ ($\Delta a_{\mu}$), which has been realized mostly with the enhancing effects in previous studies. Considering the previous discussions in our work; however, running the sparticles and Higgs boson to heavier masses, there was a possibility for that the NH contributions may reduce the SUSY contributions to muon $g-2$. After the recent updates on experimental \cite{Muong-2:2025xyk} and theoretical \cite{Aliberti:2025beg} analyses, such a suppression might be favored. Despite the slight change from the experimental measurements, the theoretical results have provided a significant improvement by including the lattice calculations which almost remove the need for BSM contributions to muon $g-2$. The new results fit with the experimental measurements as $\Delta a_{\mu} \equiv \Delta a_{\mu}^{{\rm exp}}-\Delta a_{\mu}^{{\rm SM}} \simeq 3.8(6.6)\times 10^{-10}$, which can provide a significant negative impact in SUSY models when the smuons and Bino happen to be lighter than about 1 TeV in the spectrum. 

Moreover, one of the distinctive feature of the models in the GMSB-ADJ class is to accommodate large SUSY contributions to muon $g-2$ \cite{Gogoladze:2016jvm}. In this context, negative NH contributions to muon $g-2$ can receive special importance to fit the muon $g-2$ results with the latest measurements and SM calculations. Figure \ref{fig:NH_muong2} displays our results for muon $g-2$ and the NH contributions to it in the $\Delta a_{\mu}-\Delta a_{\mu}^{H}$, $\Delta a_{\mu}-\tan\beta$, $\Delta(\Delta a_{\mu})-A_{\mu}^{\prime}$, and $m_{\tilde{\mu}_{1}}-M_{\tilde{B}}$ planes. The color coding is the same as Figure \ref{fig:L5MMess_staus}. The horizontal and vertical dashed lines indicate the bound $\Delta a_{\mu} \leq 6.6\times 10^{10}$ as recently reported by the White Paper (WP25) \cite{Aliberti:2025beg}. The horizontal line in the $\Delta a_{\mu}-\Delta a_{\mu}^{H}$ plane represents the solutions with negligible NH contributions to $\Delta a_{\mu}$ ($\Delta a_{\mu} = \Delta a_{\mu}^{H}$). Also, the diagonal line in the $m_{\tilde{\mu}_{1}}-M_{\tilde{B}}$ plane indicates the mass degeneracy between smuon and Bino ($m_{\tilde{\mu}_{1}} = M_{\tilde{B}}$). Note that we present the only solutions which yield physical (non-tachyonic) masses in both the NH and holomorphic cases, so the results show the NH contributions in the parameter space of the holomorphic GMSB-ADJ. As seen from the $\Delta a_{\mu} - \Delta a_{\mu}^{H}$ plane, one can fit the muon $g-2$ results with the NH contributions (green solutions under the horizontal dashed line), even if the holomorphic case leads to large SUSY contributions ($\Delta a_{\mu}^{H} \approx 40\times 10^{-10}$). This results allow one to fit large $\tan\beta$ solutions as well, as shown in the $\Delta a_{\mu}-\tan\beta$ plane. These negative NH contributions can be investigated in terms of the NH trilinear coupling between the smuons and $H_{u}$ denoted by $A_{\mu}^{\prime}$. The $\Delta (\Delta a_{\mu})-A_{\mu}^{\prime}$ plane shows that large $A_{\mu}^{\prime}$ values can reduce the SUSY contributions to muon $g-2$ significantly by about $50\times 10^{-10}$. This NH term can manage reducing the muon $g-2$ contributions by driving the smuon to heavier masses, but especially when it is negative, it also reduces the enhancement in SUSY contributions from the mixing in smuon by modifying it as $\mu\tan\beta \rightarrow (\mu+A_{\mu}^{\prime})\tan\beta$. With this reduction in the smuon mixing, one can still accommodate relatively light smuons and Binos in the spectrum consistently with the latest measurements and calculations of $\Delta a_{\mu}$. The allowed mass ranges for smuons and Binos are displayed in the $m_{\tilde{\mu}_{1}}-M_{\tilde{B}}$ plane of Figure \ref{fig:NH_muong2}.

\begin{table}[h!]
\centering
\setstretch{1.8}\scalebox{0.8}{\begin{tabular}{|c|ccccc|}
\hline  & Point 1 & Point 2 & Point 3 & Point 4 & Point 5 \\ \hline
$\Lambda_{5}$ &  $ 4.75 \times 10^{5} $  &  $ 2.13 \times 10^{6} $  &  $ 2.03 \times 10^{6} $  &  $ 7.44 \times 10^{6} $  &  $ 1.59 \times 10^{5} $  \\
$\Lambda_{3}$ &  $ 8.05 \times 10^{5} $  &  $ 4.89 \times 10^{5} $  &  $ 2.83 \times 10^{4} $  &  $ 4.87 \times 10^{3} $  &  $ 2.57 \times 10^{5} $  \\
$\Lambda_{8}$ &  $ 1.31 \times 10^{6} $  &  $ 2.45 \times 10^{6} $  &  $ 5.39 \times 10^{6} $  &  $ 1.45 \times 10^{4} $  &  $ 6.86 \times 10^{5} $  \\
$M_{{\rm Mess}}$ &  $ 2.73 \times 10^{8} $  &  $ 6.68 \times 10^{9} $  &  $ 1.66 \times 10^{9} $  &  $ 7.02 \times 10^{8} $  &  $ 8.17 \times 10^{7} $  \\
$\tan\beta$ &  $ 55.6 $  &  $ 51.4 $  &  $ 47.0 $  &  $ 59.9 $  &  $ 14.1 $  \\
$A_{0}^{\prime}$, $\mu^{\prime}$ &  $ 2.47 \times 10^{4} $  &  $ 3.85 \times 10^{3} $  &  $ -3.34 \times 10^{4} $  &  $ 7.93 \times 10^{4} $  &  $ 1.49 \times 10^{4} $  \\ \hline
$A_{\tau}^{\prime}$ &  $ 1.65 \times 10^{4} $  &  $ 2.27 \times 10^{3} $  &  $ -2.19 \times 10^{4} $  &  $ 7.45 \times 10^{4} $  &  $ 1.05 \times 10^{4} $  \\
$A_{b}^{\prime}$ &  $ 1.18 \times 10^{4} $  &  $ 1.44 \times 10^{3} $  &  $ -1.57 \times 10^{4} $  &  $ 7.45 \times 10^{4} $  &  $ 7.68 \times 10^{3} $  \\
$A_{t}^{\prime}$ &  $ 1.49 \times 10^{4} $  &  $ 1.88 \times 10^{3} $  &  $ -1.90 \times 10^{4} $  &  $ 7.45 \times 10^{4} $  &  $ 1.04 \times 10^{4} $  \\ \hline
$m_{\tilde{\tau}_{1}}$, $\Delta m_{\tilde{\tau}_{1}}$ &  $ \cred{5994} $ ,  $ \cred{5931} $  &  $ \cred{3590} $ ,  $ \cred{-274.8} $  &  $ 10131 $ ,  $ 6446 $  &  $ 23869 $ ,  $ 13880 $  &  $ 1027 $ ,  $ 772.3 $  \\
$m_{\tilde{b}_{1}}$, $\Delta m_{\tilde{b}_{1}}$ &  $ 22152 $ ,  $ 1192 $  &  $ 41871 $ ,  $ \cblue{20.5} $  &  $ 78885 $ ,  $ 905.3 $  &  $ 53906 $ ,  $ 4566 $  &  $ 11917 $ ,  $ 448.4 $  \\
$m_{\tilde{t}_{1}}$, $\Delta m_{\tilde{t}_{1}}$ &  $ 22088 $ ,  $ 1915 $  &  $ 40082 $ ,  $ \cblue{-23.2} $  &  $ 77200 $ ,  $ 1561 $  &  $ 54284 $ ,  $ 5620 $  &  $ 12101 $ ,  $ 1241 $  \\
$M_{{\rm SUSY}}$, $\Delta M_{{\rm SUSY}}$ &  $ 22331 $ ,  $ 2229 $  &  $ 40968 $ ,  $ 971.9 $  &  $ 78038 $ ,  $ 3016 $  &  $ 56526 $ ,  $ 5640 $  &  $ 12135 $ ,  $ 1284 $  \\ \hline
$m_{h}$, $\Delta m_{h}$ &  $ 124.6 $ ,  $ 17.0 $  &  $ 123.5 $ ,  $ -2.0 $  &  $ \cred{125.6} $ ,  $ \cred{52.8} $  &  $ \cred{125.6} $ ,  $ \cred{-3.7} $  &  $ 124.1 $ ,  $ -5.4 $  \\
$\Delta a_{\mu}\times 10^{10}$, $\Delta (\Delta a_{\mu})\times 10^{10}$ &  $ 1.2 $ ,  $ -0.1 $  &  $ 0.2 $ ,  $ 0.0 $  &  $ 0.3 $ ,  $ -0.2 $  &  $ 0.0 $ ,  $ 0.0 $  &  $ \cred{3.0} $ ,  $ \cred{-22.1} $  \\ \hline
\end{tabular}
}
\caption{Benchmark points exemplifying our findings. All points are selected to be consistent with the experimental constraints employed in our analyses. All the points are selected by requiring to be consistent with the experimental constraints employed in our analyses. All the parameters, except $\Delta a_{\mu}$, are expressed in units of GeV. The red color emphasize the parameters relevant to our discussions, while the blue color shows the cases with negligible NH contributions.}
\label{tab1:Bench}
\end{table}

Before we conclude, we exemplify our findings with several benchmark scenarios in Table \ref{tab1:Bench}. All the points are selected by requiring to be consistent with the experimental constraints employed in our analyses. All the parameters, except $\Delta a_{\mu}$, are expressed in units of GeV. Point 1 depicts a solution which involves inconsistently light staus ($\simeq 60$ GeV) in the spectrum without the NH contributions. The NH terms are induced at the order of $10^{4}$ GeV in such solutions, and their contributions shift the stau mass to about 6 TeV. The sbottom and stop receive about 1-2 TeV NH contributions. The SM-like Higgs boson mass is also lifted by about 17 GeV and becomes consistent with the experimental constraints. In contrast to Point1, Point 2 exemplifies the largest negative NH contributions to the stau mass which is realized in our analyses at about 275 GeV. These solutions represented by Point 2 induce the NH terms smaller by one order of magnitude in comparison with those displayed by Point 1. In addition, the $M_{{\rm Mess}}$ happens one order of magnitude larger. The contributions to sbottom and stop are about 20 GeV, and they are negligible compared to their masses, though the NH contributions modifies the renormalization scale ($M_{{\rm SUSY}}$) by about 1 TeV, which is also negligible comparing to $M_{{\rm SUSY}}$. The negative contributions to stau and stop in these solutions leads to about 2 GeV negative contribution in the SM-like Higgs Boson mass. Points 3 and 4 represent the SM-like Higgs boson solutions which coincide with its experimental value ($\sim 125.6$ GeV). Point 3 display a solution in which the SM-like Higgs boson receives relatively large and positive NH contributions, and its mass is shifted from about 70 GeV to 125.6 GeV. On the other hand, Point 4 shows negative contributions to realize 125.6 GeV Higgs boson. Such contributions are realized at about -4 GeV in our analyses. All the points except Point 4 exhibit negligible contributions to muon $g-2$ due to the heavy spectrum. Point 5, on the other hand, depicts a solution which predicts a large muon $g-2$ results in the holomorphic case, while the result fits with the current constraint on muon $g-2$ after receiving the NH contributions.

\section{Conclusion}
\label{sec:conc}

We explore the implications of the models in which the SUSY breaking is transmitted to the visible sector through the gauge interactions. In this class of models, the messenger fields reside in the adjoint representation of the MSSM gauge group, and considering the complete representation the fields do not lead to non-zero anomalies. We include the NH terms in the Lagrangian, which are induced at the third-loop level of the perturbation theory. In our case, there is no SUSY breaking singlet field, so the infinities and the hierarchy problem in the SM-like Higgs boson mass may not arise. In the framework of the GMSB-ADJ, we link such NH terms to the SUSY breaking scales, which brings a further suppression on them by $1/M_{{\rm Mess}}$. Despite this further suppression, we observe that such NH terms can be as large as about 200 TeV, when the SUSY breaking in the hidden sector happens at a relatively large scales, while the visible sector sees it at lower scales ($M_{{\rm Mess}} \lesssim 10^{11}$ GeV). Previous studies consider such terms rather in an effective approach in which they contribute to the physical observables through the mixing of the sparticles. These contributions can significantly alter the masses especially for the stau and sbottom, since they are enhanced with $\tan\beta$, while they provide minor changes in stops due to the suppression by $\tan\beta$. Besides, the Higgsino masses can be realized as $\mu + \mu^{\prime}$, where $\mu^{\prime}$ is induced NH Higgsino mass. Therefore, one can fit heavy Higgsino masses even in natural SUSY models. Apart from raising the masses of the sparticles and SM-like Higgs boson, they provide also significant contributions to muon $g-2$, which may receive a strong negative impact from the latest reports about its experimental measurements and theoretical calculations.

In our work, we impose the NH terms at the SUSY breaking scale in the visible sector by linking them numerically to the masses of the messengers as $A_{0}^{\prime} \sim \Lambda_{i}^{2}/M_{{\rm Mess}}$, and include them in RG evolutions of the SSB parameters. The main impact, especially those from their RG contributions, is that the parameter space of the models is significantly enlarged. The small $\Lambda_{5}$ values mostly lead to tachyonic right-handed staus without the NH contributions, while even small NH terms can keep its mass-square positive through the RG evolutions. It allows one to explore the regions with small $\Lambda_{5}$. We observe that the NH contributions can differ the right-handed stau mass by about 25 TeV. This difference reduces to about 20 TeV at most when one includes their effect in stau mixing in calculating the mass eigenstates of staus. Similarly, these terms can yield heavier stop at the low scale by about 15 TeV, and such a large difference in stop masses can yield significant enhancement in the Higgs boson mass by about 80 GeV, and one can realize consistent masses for the SM-like Higgs boson, even they are very light without the NH contributions. Despite expecting large contributions for sbottoms, we realize only about 6-7 TeV difference in sbottom mass by comparing with the holomorphic case. We find that the NH contributions through the RGEs and sparticle mixing rather cancel each other. Even though we observe mostly positive contributions from the NH terms, in a small region these contributions can slightly lower the mass spectrum by about 1 TeV. In these regions, also the SM-like Higgs mass decreases about 20 GeV. 

An interesting effect from the NH contributions can be seen in the SUSY contributions to muon $g-2$. The latest measurements strengthen the consensus that the SM may not need BSM contributions to fit its predictions on muon $g-2$ with the experimental measurements. A negative impact from the latest status of muon $g-2$ may exclude the light slepton and gaugino solutions. One can expect that the NH contributions suppress the SUSY contributions to muon $g-2$, since they drive the SUSY particles heavier masses in most of the parameter space. However, this observation still excludes the light sparticles. On the other hand, the NH terms do not only alter the muon $g-2$ results by raising the sparticle masses. They directly contribute to the muon $g-2$ through smuon mixing, and these contributions are enhanced by $\tan\beta$. In our analyses, we realize mostly negative contributions to muon $g-2$ in the experimentally consistent regions (green in the plots), and they can even change the muon $g-2$ results by about $-50\times 10^{-10}$. In this context, the large muon $g-2$ solutions in the holomorphic case can be fit with the current measurements and calculations after including the NH contributions. Consequently, the light sparticle and gaugino solutions can still survive. 

\noindent {\bf Acknowledgement:}

The work represented in this study is supported by the Scientific and Technological Research Council of Turkey (TUBITAK) Grant. No. MFAG-124F280. The numerical calculations reported in this paper were partially performed at TUBITAK ULAKBIM, High Performance and Grid Computing Center (TRUBA resources).

\providecommand{\href}[2]{#2}\begingroup\raggedright\endgroup


\begin{thebibliography}{10}

\bibitem{ATLAS:2012yve}
{\scshape ATLAS} collaboration, \emph{{Observation of a new particle in the
  search for the Standard Model Higgs boson with the ATLAS detector at the
  LHC}}, \href{https://doi.org/10.1016/j.physletb.2012.08.020}{\emph{Phys.
  Lett. B} {\bfseries 716} (2012) 1}
  [\href{https://arxiv.org/abs/1207.7214}{{\ttfamily 1207.7214}}].

\bibitem{CMS:2012qbp}
{\scshape CMS} collaboration, \emph{{Observation of a New Boson at a Mass of
  125 GeV with the CMS Experiment at the LHC}},
  \href{https://doi.org/10.1016/j.physletb.2012.08.021}{\emph{Phys. Lett. B}
  {\bfseries 716} (2012) 30} [\href{https://arxiv.org/abs/1207.7235}{{\ttfamily
  1207.7235}}].

\bibitem{CMS:2013btf}
{\scshape CMS} collaboration, \emph{{Observation of a New Boson with Mass Near
  125 GeV in $pp$ Collisions at $\sqrt{s}$ = 7 and 8 TeV}},
  \href{https://doi.org/10.1007/JHEP06(2013)081}{\emph{JHEP} {\bfseries 06}
  (2013) 081} [\href{https://arxiv.org/abs/1303.4571}{{\ttfamily 1303.4571}}].

\bibitem{Degrassi:2012ry}
G.~Degrassi, S.~Di~Vita, J.~Elias-Miro, J.R.~Espinosa, G.F.~Giudice, G.~Isidori
  et~al., \emph{{Higgs mass and vacuum stability in the Standard Model at
  NNLO}}, \href{https://doi.org/10.1007/JHEP08(2012)098}{\emph{JHEP} {\bfseries
  08} (2012) 098} [\href{https://arxiv.org/abs/1205.6497}{{\ttfamily
  1205.6497}}].

\bibitem{Bezrukov:2012sa}
F.~Bezrukov, M.Y.~Kalmykov, B.A.~Kniehl and M.~Shaposhnikov, \emph{{Higgs Boson
  Mass and New Physics}},
  \href{https://doi.org/10.1007/JHEP10(2012)140}{\emph{JHEP} {\bfseries 10}
  (2012) 140} [\href{https://arxiv.org/abs/1205.2893}{{\ttfamily 1205.2893}}].

\bibitem{Buttazzo:2013uya}
D.~Buttazzo, G.~Degrassi, P.P.~Giardino, G.F.~Giudice, F.~Sala, A.~Salvio
  et~al., \emph{{Investigating the near-criticality of the Higgs boson}},
  \href{https://doi.org/10.1007/JHEP12(2013)089}{\emph{JHEP} {\bfseries 12}
  (2013) 089} [\href{https://arxiv.org/abs/1307.3536}{{\ttfamily 1307.3536}}].

\bibitem{Carena:2012mw}
M.~Carena, S.~Gori, I.~Low, N.R.~Shah and C.E.M.~Wagner, \emph{{Vacuum
  Stability and Higgs Diphoton Decays in the MSSM}},
  \href{https://doi.org/10.1007/JHEP02(2013)114}{\emph{JHEP} {\bfseries 02}
  (2013) 114} [\href{https://arxiv.org/abs/1211.6136}{{\ttfamily 1211.6136}}].

\bibitem{Carena:2011aa}
M.~Carena, S.~Gori, N.R.~Shah and C.E.M.~Wagner, \emph{{A 125 GeV SM-like Higgs
  in the MSSM and the $\gamma \gamma$ rate}},
  \href{https://doi.org/10.1007/JHEP03(2012)014}{\emph{JHEP} {\bfseries 03}
  (2012) 014} [\href{https://arxiv.org/abs/1112.3336}{{\ttfamily 1112.3336}}].

\bibitem{Ajaib:2012vc}
M.A.~Ajaib, I.~Gogoladze, F.~Nasir and Q.~Shafi, \emph{{Revisiting mGMSB in
  Light of a 125 GeV Higgs}},
  \href{https://doi.org/10.1016/j.physletb.2012.06.036}{\emph{Phys. Lett. B}
  {\bfseries 713} (2012) 462}
  [\href{https://arxiv.org/abs/1204.2856}{{\ttfamily 1204.2856}}].

\bibitem{Gogoladze:2015tfa}
I.~Gogoladze, A.~Mustafayev, Q.~Shafi and C.S.~Un, \emph{{Yukawa Unification
  and Sparticle Spectroscopy in Gauge Mediation Models}},
  \href{https://doi.org/10.1103/PhysRevD.91.096005}{\emph{Phys. Rev. D}
  {\bfseries 91} (2015) 096005}
  [\href{https://arxiv.org/abs/1501.07290}{{\ttfamily 1501.07290}}].

\bibitem{ParticleDataGroup:2024cfk}
{\scshape Particle Data Group} collaboration, \emph{{Review of particle
  physics}}, \href{https://doi.org/10.1103/PhysRevD.110.030001}{\emph{Phys.
  Rev. D} {\bfseries 110} (2024) 030001}.

\bibitem{Goodsell:2014bna}
M.D.~Goodsell, K.~Nickel and F.~Staub, \emph{{Two-Loop Higgs mass calculations
  in supersymmetric models beyond the MSSM with SARAH and SPheno}},
  \href{https://doi.org/10.1140/epjc/s10052-014-3247-y}{\emph{Eur. Phys. J. C}
  {\bfseries 75} (2015) 32} [\href{https://arxiv.org/abs/1411.0675}{{\ttfamily
  1411.0675}}].

\bibitem{Allanach:2001kg}
B.C.~Allanach, \emph{{SOFTSUSY: a program for calculating supersymmetric
  spectra}}, \href{https://doi.org/10.1016/S0010-4655(01)00460-X}{\emph{Comput.
  Phys. Commun.} {\bfseries 143} (2002) 305}
  [\href{https://arxiv.org/abs/hep-ph/0104145}{{\ttfamily hep-ph/0104145}}].

\bibitem{Allanach:2014nba}
B.C.~Allanach, A.~Bednyakov and R.~Ruiz~de Austri, \emph{{Higher order
  corrections and unification in the minimal supersymmetric standard model:
  SOFTSUSY3.5}}, \href{https://doi.org/10.1016/j.cpc.2014.12.006}{\emph{Comput.
  Phys. Commun.} {\bfseries 189} (2015) 192}
  [\href{https://arxiv.org/abs/1407.6130}{{\ttfamily 1407.6130}}].

\bibitem{Baer:2021tta}
H.~Baer, V.~Barger and D.~Martinez, \emph{{Comparison of SUSY spectra
  generators for natural SUSY and string landscape predictions}},
  \href{https://doi.org/10.1140/epjc/s10052-022-10141-2}{\emph{Eur. Phys. J. C}
  {\bfseries 82} (2022) 172}
  [\href{https://arxiv.org/abs/2111.03096}{{\ttfamily 2111.03096}}].

\bibitem{Gomez:2022qrb}
M.E.~Gomez, Q.~Shafi, A.~Tiwari and C.S.~Un, \emph{{Muon $\mathbf {g-2}$,
  neutralino dark matter and stau NLSP}},
  \href{https://doi.org/10.1140/epjc/s10052-022-10507-6}{\emph{Eur. Phys. J. C}
  {\bfseries 82} (2022) 561}
  [\href{https://arxiv.org/abs/2202.06419}{{\ttfamily 2202.06419}}].

\bibitem{Gogoladze:2015jua}
I.~Gogoladze, Q.~Shafi and C.S.~\"Un, \emph{{Reconciling the muon
  g\ensuremath{-}2 , a 125 GeV Higgs boson, and dark matter in gauge mediation
  models}}, \href{https://doi.org/10.1103/PhysRevD.92.115014}{\emph{Phys. Rev.
  D} {\bfseries 92} (2015) 115014}
  [\href{https://arxiv.org/abs/1509.07906}{{\ttfamily 1509.07906}}].

\bibitem{Han:1998pa}
T.~Han, T.~Yanagida and R.-J.~Zhang, \emph{{Adjoint messengers and perturbative
  unification at the string scale}},
  \href{https://doi.org/10.1103/PhysRevD.58.095011}{\emph{Phys. Rev. D}
  {\bfseries 58} (1998) 095011}
  [\href{https://arxiv.org/abs/hep-ph/9804228}{{\ttfamily hep-ph/9804228}}].

\bibitem{Bhattacharyya:2013xma}
G.~Bhattacharyya, B.~Bhattacherjee, T.T.~Yanagida and N.~Yokozaki, \emph{{A
  practical GMSB model for explaining the muon (g-2) with gauge coupling
  unification}},
  \href{https://doi.org/10.1016/j.physletb.2013.12.064}{\emph{Phys. Lett. B}
  {\bfseries 730} (2014) 231}
  [\href{https://arxiv.org/abs/1311.1906}{{\ttfamily 1311.1906}}].

\bibitem{Bhattacharyya:2015vha}
G.~Bhattacharyya, T.T.~Yanagida and N.~Yokozaki, \emph{{Focus Point Gauge
  Mediation with Incomplete Adjoint Messengers and Gauge Coupling
  Unification}},
  \href{https://doi.org/10.1016/j.physletb.2015.07.052}{\emph{Phys. Lett. B}
  {\bfseries 749} (2015) 82}
  [\href{https://arxiv.org/abs/1506.05962}{{\ttfamily 1506.05962}}].

\bibitem{Gogoladze:2016grr}
I.~Gogoladze, A.~Mustafayev, Q.~Shafi and C.S.~Un, \emph{{Gauge Mediation
  Models with Adjoint Messengers}},
  \href{https://doi.org/10.1103/PhysRevD.94.075012}{\emph{Phys. Rev. D}
  {\bfseries 94} (2016) 075012}
  [\href{https://arxiv.org/abs/1609.02124}{{\ttfamily 1609.02124}}].

\bibitem{Gogoladze:2016jvm}
I.~Gogoladze and C.S.~Un, \emph{{Muon g - 2 in gauge mediated supersymmetry
  breaking models with adjoint messengers}},
  \href{https://doi.org/10.1103/PhysRevD.95.035028}{\emph{Phys. Rev. D}
  {\bfseries 95} (2017) 035028}
  [\href{https://arxiv.org/abs/1612.02376}{{\ttfamily 1612.02376}}].

\bibitem{Inoue:1982pi}
K.~Inoue, A.~Kakuto, H.~Komatsu and S.~Takeshita, \emph{{Aspects of Grand
  Unified Models with Softly Broken Supersymmetry}},
  \href{https://doi.org/10.1143/PTP.68.927}{\emph{Prog. Theor. Phys.}
  {\bfseries 68} (1982) 927}.

\bibitem{Hall:1990ac}
L.J.~Hall and L.~Randall, \emph{{Weak scale effective supersymmetry}},
  \href{https://doi.org/10.1103/PhysRevLett.65.2939}{\emph{Phys. Rev. Lett.}
  {\bfseries 65} (1990) 2939}.

\bibitem{Bagger:1993ji}
J.~Bagger and E.~Poppitz, \emph{{Destabilizing divergences in supergravity
  coupled supersymmetric theories}},
  \href{https://doi.org/10.1103/PhysRevLett.71.2380}{\emph{Phys. Rev. Lett.}
  {\bfseries 71} (1993) 2380}
  [\href{https://arxiv.org/abs/hep-ph/9307317}{{\ttfamily hep-ph/9307317}}].

\bibitem{Bagger:1995ay}
J.~Bagger, E.~Poppitz and L.~Randall, \emph{{Destabilizing divergences in
  supergravity theories at two loops}},
  \href{https://doi.org/10.1016/0550-3213(95)00463-3}{\emph{Nucl. Phys. B}
  {\bfseries 455} (1995) 59}
  [\href{https://arxiv.org/abs/hep-ph/9505244}{{\ttfamily hep-ph/9505244}}].

\bibitem{Martin:1999hc}
S.P.~Martin, \emph{{Dimensionless supersymmetry breaking couplings, flat
  directions, and the origin of intermediate mass scales}},
  \href{https://doi.org/10.1103/PhysRevD.61.035004}{\emph{Phys. Rev. D}
  {\bfseries 61} (2000) 035004}
  [\href{https://arxiv.org/abs/hep-ph/9907550}{{\ttfamily hep-ph/9907550}}].

\bibitem{Jack:1999ud}
I.~Jack and D.R.T.~Jones, \emph{{Nonstandard soft supersymmetry breaking}},
  \href{https://doi.org/10.1016/S0370-2693(99)00530-4}{\emph{Phys. Lett. B}
  {\bfseries 457} (1999) 101}
  [\href{https://arxiv.org/abs/hep-ph/9903365}{{\ttfamily hep-ph/9903365}}].

\bibitem{Jack:1999fa}
I.~Jack and D.R.T.~Jones, \emph{{Quasiinfrared fixed points and renormalization
  group invariant trajectories for nonholomorphic soft supersymmetry
  breaking}}, \href{https://doi.org/10.1103/PhysRevD.61.095002}{\emph{Phys.
  Rev. D} {\bfseries 61} (2000) 095002}
  [\href{https://arxiv.org/abs/hep-ph/9909570}{{\ttfamily hep-ph/9909570}}].

\bibitem{Haber:2007dj}
H.E.~Haber and J.D.~Mason, \emph{{Hard supersymmetry-breaking 'wrong-Higgs'
  couplings of the MSSM}},
  \href{https://doi.org/10.1103/PhysRevD.77.115011}{\emph{Phys. Rev. D}
  {\bfseries 77} (2008) 115011}
  [\href{https://arxiv.org/abs/0711.2890}{{\ttfamily 0711.2890}}].

\bibitem{Chattopadhyay:2017qvh}
U.~Chattopadhyay, D.~Das and S.~Mukherjee, \emph{{Exploring Non-Holomorphic
  Soft Terms in the Framework of Gauge Mediated Supersymmetry Breaking}},
  \href{https://doi.org/10.1007/JHEP01(2018)158}{\emph{JHEP} {\bfseries 01}
  (2018) 158} [\href{https://arxiv.org/abs/1710.10120}{{\ttfamily
  1710.10120}}].

\bibitem{Ali:2021kxa}
M.I.~Ali, M.~Chakraborti, U.~Chattopadhyay and S.~Mukherjee, \emph{{Muon and
  electron $(g-2)$ anomalies with non-holomorphic interactions in MSSM}},
  \href{https://doi.org/10.1140/epjc/s10052-023-11216-4}{\emph{Eur. Phys. J. C}
  {\bfseries 83} (2023) 60} [\href{https://arxiv.org/abs/2112.09867}{{\ttfamily
  2112.09867}}].

\bibitem{Chakraborty:2019wav}
S.~Chakraborty and T.S.~Roy, \emph{{Radiatively generated source of flavor
  universal scalar soft masses}},
  \href{https://doi.org/10.1103/PhysRevD.100.035020}{\emph{Phys. Rev. D}
  {\bfseries 100} (2019) 035020}
  [\href{https://arxiv.org/abs/1904.10144}{{\ttfamily 1904.10144}}].

\bibitem{Un:2014afa}
C.S.~\"Un, c.H.~Tany\i{}ld\i{}z\i{}, S.~Kerman and L.~Solmaz,
  \emph{{Generalized Soft Breaking Leverage for the MSSM}},
  \href{https://doi.org/10.1103/PhysRevD.91.105033}{\emph{Phys. Rev. D}
  {\bfseries 91} (2015) 105033}
  [\href{https://arxiv.org/abs/1412.1440}{{\ttfamily 1412.1440}}].

\bibitem{Rehman:2022ydc}
M.~Rehman and S.~Heinemeyer, \emph{{Nonholomorphic soft-term contributions to
  the Higgs-boson masses in the Feynman diagrammatic approach}},
  \href{https://doi.org/10.1103/PhysRevD.107.095033}{\emph{Phys. Rev. D}
  {\bfseries 107} (2023) 095033}
  [\href{https://arxiv.org/abs/2212.13757}{{\ttfamily 2212.13757}}].

\bibitem{Un:2023wws}
C.S.~Un, \emph{{Low fine-tuning with heavy higgsinos in Yukawa unified SUSY
  GUTs}}, \href{https://doi.org/10.55730/1300-0101.2753}{\emph{Turk. J. Phys.}
  {\bfseries 48} (2024) 1} [\href{https://arxiv.org/abs/2308.12862}{{\ttfamily
  2308.12862}}].

\bibitem{Israr:2025cfd}
S.~Israr, M.E.~G\'omez and M.~Rehman, \emph{{Nonholomorphic Higgsino Mass Term
  Effects on Muon g \ensuremath{-} 2 and Dark Matter Relic Density in Flavor
  Symmetry-Based Minimal Supersymmetric Standard Model}},
  \href{https://doi.org/10.3390/particles8010030}{\emph{Particles} {\bfseries
  8} (2025) 30}.

\bibitem{Israr:2024ubp}
S.~Israr and M.~Rehman, \emph{{Higgs decay to $Z\gamma $ in the minimal
  supersymmetric standard model and its nonholomorphic extension}},
  \href{https://doi.org/10.1140/epjp/s13360-025-06401-1}{\emph{Eur. Phys. J.
  Plus} {\bfseries 140} (2025) 397}
  [\href{https://arxiv.org/abs/2407.01210}{{\ttfamily 2407.01210}}].

\bibitem{Rehman:2025djc}
M.~Rehman and S.~Heinemeyer, \emph{{Toward a refined understanding of
  nonholomorphic soft SUSY-breaking effects on the Higgs boson mass spectra}},
  \href{https://doi.org/10.1103/n5tq-1h1y}{\emph{Phys. Rev. D} {\bfseries 111}
  (2025) 115027} [\href{https://arxiv.org/abs/2504.11891}{{\ttfamily
  2504.11891}}].

\bibitem{Hamaguchi:2014sea}
K.~Hamaguchi, M.~Ibe, T.T.~Yanagida and N.~Yokozaki, \emph{{Testing the Minimal
  Direct Gauge Mediation at the LHC}},
  \href{https://doi.org/10.1103/PhysRevD.90.015027}{\emph{Phys. Rev. D}
  {\bfseries 90} (2014) 015027}
  [\href{https://arxiv.org/abs/1403.1398}{{\ttfamily 1403.1398}}].

\bibitem{Delgado:2013gza}
A.~Delgado, M.~Garcia and M.~Quiros, \emph{{Electroweak and supersymmetry
  breaking from the Higgs boson discovery}},
  \href{https://doi.org/10.1103/PhysRevD.90.015016}{\emph{Phys. Rev. D}
  {\bfseries 90} (2014) 015016}
  [\href{https://arxiv.org/abs/1312.3235}{{\ttfamily 1312.3235}}].

\bibitem{Draper:2011aa}
P.~Draper, P.~Meade, M.~Reece and D.~Shih, \emph{{Implications of a 125 GeV
  Higgs for the MSSM and Low-Scale SUSY Breaking}},
  \href{https://doi.org/10.1103/PhysRevD.85.095007}{\emph{Phys. Rev. D}
  {\bfseries 85} (2012) 095007}
  [\href{https://arxiv.org/abs/1112.3068}{{\ttfamily 1112.3068}}].

\bibitem{Draper:2013oza}
P.~Draper, G.~Lee and C.E.M.~Wagner, \emph{{Precise estimates of the Higgs mass
  in heavy supersymmetry}},
  \href{https://doi.org/10.1103/PhysRevD.89.055023}{\emph{Phys. Rev. D}
  {\bfseries 89} (2014) 055023}
  [\href{https://arxiv.org/abs/1312.5743}{{\ttfamily 1312.5743}}].

\bibitem{Giudice:1998bp}
G.F.~Giudice and R.~Rattazzi, \emph{{Theories with gauge mediated supersymmetry
  breaking}}, \href{https://doi.org/10.1016/S0370-1573(99)00042-3}{\emph{Phys.
  Rept.} {\bfseries 322} (1999) 419}
  [\href{https://arxiv.org/abs/hep-ph/9801271}{{\ttfamily hep-ph/9801271}}].

\bibitem{McGuirk:2012sb}
P.~McGuirk, G.~Shiu and F.~Ye, \emph{{Soft branes in supersymmetry-breaking
  backgrounds}}, \href{https://doi.org/10.1007/JHEP07(2012)188}{\emph{JHEP}
  {\bfseries 07} (2012) 188} [\href{https://arxiv.org/abs/1206.0754}{{\ttfamily
  1206.0754}}].

\bibitem{Gogoladze:2012ii}
I.~Gogoladze, Q.~Shafi and C.S.~Un, \emph{{125 GeV Higgs Boson from t-b-tau
  Yukawa Unification}},
  \href{https://doi.org/10.1007/JHEP07(2012)055}{\emph{JHEP} {\bfseries 07}
  (2012) 055} [\href{https://arxiv.org/abs/1203.6082}{{\ttfamily 1203.6082}}].

\bibitem{Martin:1997ns}
S.P.~Martin, \emph{{A Supersymmetry primer}},
  \href{https://doi.org/10.1142/9789812839657_0001}{\emph{Adv. Ser. Direct.
  High Energy Phys.} {\bfseries 18} (1998) 1}
  [\href{https://arxiv.org/abs/hep-ph/9709356}{{\ttfamily hep-ph/9709356}}].

\bibitem{Kitahara:2013lfa}
T.~Kitahara and T.~Yoshinaga, \emph{{Stau with Large Mass Difference and
  Enhancement of the Higgs to Diphoton Decay Rate in the MSSM}},
  \href{https://doi.org/10.1007/JHEP05(2013)035}{\emph{JHEP} {\bfseries 05}
  (2013) 035} [\href{https://arxiv.org/abs/1303.0461}{{\ttfamily 1303.0461}}].

\bibitem{Bellisai:1997ck}
D.~Bellisai, F.~Fucito, M.~Matone and G.~Travaglini, \emph{{Nonholomorphic
  terms in N=2 SUSY Wilsonian actions and the renormalization group equation}},
  \href{https://doi.org/10.1103/PhysRevD.56.5218}{\emph{Phys. Rev. D}
  {\bfseries 56} (1997) 5218}
  [\href{https://arxiv.org/abs/hep-th/9706099}{{\ttfamily hep-th/9706099}}].

\bibitem{Bergamin:2003ub}
L.~Bergamin and P.~Minkowski, \emph{{SUSY glue balls, dynamical symmetry
  breaking and nonholomorphic potentials}},
  \href{https://arxiv.org/abs/hep-th/0301155}{{\ttfamily hep-th/0301155}}.

\bibitem{Porod:2003um}
W.~Porod, \emph{{SPheno, a program for calculating supersymmetric spectra, SUSY
  particle decays and SUSY particle production at e+ e- colliders}},
  \href{https://doi.org/10.1016/S0010-4655(03)00222-4}{\emph{Comput. Phys.
  Commun.} {\bfseries 153} (2003) 275}
  [\href{https://arxiv.org/abs/hep-ph/0301101}{{\ttfamily hep-ph/0301101}}].

\bibitem{Staub:2008uz}
F.~Staub, \emph{{SARAH}}, {\emph{arXiv} {\bfseries 0806.0538} (2008) }.

\bibitem{Staub:2015iza}
F.~Staub, \emph{{Introduction to SARAH and related tools}},
  \href{https://doi.org/10.22323/1.263.0027}{\emph{PoS} {\bfseries CORFU2015}
  (2016) 027} [\href{https://arxiv.org/abs/1509.07061}{{\ttfamily
  1509.07061}}].

\bibitem{Baer:2008jn}
H.~Baer, S.~Kraml, S.~Sekmen and H.~Summy, \emph{{Dark matter allowed scenarios
  for Yukawa-unified SO(10) SUSY GUTs}},
  \href{https://doi.org/10.1088/1126-6708/2008/03/056}{\emph{JHEP} {\bfseries
  03} (2008) 056} [\href{https://arxiv.org/abs/0801.1831}{{\ttfamily
  0801.1831}}].

\bibitem{Belanger:2009ti}
G.~Belanger, F.~Boudjema, A.~Pukhov and R.K.~Singh, \emph{{Constraining the
  MSSM with universal gaugino masses and implication for searches at the LHC}},
  \href{https://doi.org/10.1088/1126-6708/2009/11/026}{\emph{JHEP} {\bfseries
  11} (2009) 026} [\href{https://arxiv.org/abs/0906.5048}{{\ttfamily
  0906.5048}}].

\bibitem{ParticleDataGroup:2014cgo}
{\scshape Particle Data Group} collaboration, \emph{{Review of Particle
  Physics}}, \href{https://doi.org/10.1088/1674-1137/38/9/090001}{\emph{Chin.
  Phys. C} {\bfseries 38} (2014) 090001}.

\bibitem{ATLAS:2021twp}
{\scshape ATLAS} collaboration, \emph{{Search for squarks and gluinos in final
  states with one isolated lepton, jets, and missing transverse momentum at
  $\sqrt{s}=13$~ with the ATLAS detector}},
  \href{https://doi.org/10.1140/epjc/s10052-021-09748-8}{\emph{Eur. Phys. J. C}
  {\bfseries 81} (2021) 600}
  [\href{https://arxiv.org/abs/2101.01629}{{\ttfamily 2101.01629}}].

\bibitem{ATLAS:2020syg}
{\scshape ATLAS} collaboration, \emph{{Search for squarks and gluinos in final
  states with jets and missing transverse momentum using 139 fb$^{-1}$ of
  $\sqrt{s}$ =13 TeV $pp$ collision data with the ATLAS detector}},
  \href{https://doi.org/10.1007/JHEP02(2021)143}{\emph{JHEP} {\bfseries 02}
  (2021) 143} [\href{https://arxiv.org/abs/2010.14293}{{\ttfamily
  2010.14293}}].

\bibitem{ATLAS:2022rcw}
{\scshape ATLAS} collaboration, \emph{{SUSY Summary Plots March 2022}},
  ATL-PHYS-PUB-2022-013.

\bibitem{Belle-II:2022hys}
{\scshape Belle-II} collaboration, \emph{{Measurement of the photon-energy
  spectrum in inclusive $B\rightarrow X_{s}\gamma$ decays identified using
  hadronic decays of the recoil $B$ meson in 2019-2021 Belle II data}},
  {\emph{BELLE2-CONF-PH-2022-018} {\bfseries 2210.10220} (2022) }.

\bibitem{CMS:2020rox}
{CMS Collaboration}, \emph{{Combination of the ATLAS, CMS and LHCb results on
  the $B^0_{(s)} \to \mu^+\mu^-$ decays}}, {\emph{CMS-PAS-BPH-20-003} (2020) }.

\bibitem{Aliberti:2025beg}
R.~Aliberti et~al., \emph{{The anomalous magnetic moment of the muon in the
  Standard Model: an update}},
  \href{https://arxiv.org/abs/2505.21476}{{\ttfamily 2505.21476}}.

\bibitem{Muong-2:2025xyk}
{\scshape Muon g-2} collaboration, \emph{{Measurement of the Positive Muon
  Anomalous Magnetic Moment to 127 ppb}},
  \href{https://arxiv.org/abs/2506.03069}{{\ttfamily 2506.03069}}.

\end{thebibliography}

\end{document}